\documentclass[aps,unsortedaddress,10pt,nobibnotes,notitlepage,twocolumn,longbibliography]{revtex4-1}

\usepackage[normalem]{ulem}

\usepackage[latin1]{inputenc}
\usepackage{graphicx}
\usepackage{float}
\usepackage{latexsym}
\usepackage{graphicx}
\usepackage{amssymb}
\usepackage{amsmath}
\usepackage{amsfonts}
\usepackage[colorlinks=true,citecolor=blue,hyperfootnotes=false]{hyperref}

\usepackage{cool}
\usepackage{dcolumn}
\usepackage{textcomp}
\usepackage[dvipsnames]{xcolor}
\usepackage{xfrac}
\usepackage{slashed}
\usepackage{multirow}
\usepackage{derivative}
\usepackage{tensor}
\usepackage{physics}
\usepackage{comment}

\def\figs/B{B}
\def\bea{\begin{eqnarray}}
\def\eea{\end{eqnarray}}
\def\bg{\begin{eqnarray}}
\def\nd{\end{eqnarray}}

\def\beq{\begin{equation}}
\def\eeq{\end{equation}}

\begin{document}

\title{Primordial Black Holes with QCD Color Charge}

\author{Elba Alonso-Monsalve}
\email{elba\_am@mit.edu}

\author{David I.~Kaiser}
\email{dikaiser@mit.edu}

\affiliation{Department of Physics, Massachusetts Institute of Technology, Cambridge, MA 02139, USA}

\begin{abstract}
    We describe a realistic mechanism whereby black holes with significant QCD color charge could have formed during the early universe. Primordial black holes (PBHs) could make up a significant fraction of the dark matter if they formed well before the QCD confinement transition. Such PBHs would form by absorbing unconfined quarks and gluons, and hence could acquire a net color charge. We estimate the number of PBHs per Hubble volume with near-extremal color charge for various scenarios, and discuss possible phenomenological implications.
\end{abstract}

\maketitle

{\it Introduction.} Primordial black holes (PBHs) were proposed more than half a century ago \cite{Zeldovich:1967lct,Hawking:1971ei,Carr:1974nx}, and they remain a fascinating theoretical curiosity. In Hawking's first paper on PBHs he even suggested they might be relevant to the ``missing mass'' puzzle identified by astronomers \cite{Hawking:1971ei}. In recent years, the possibility that PBHs might constitute a significant fraction of the present-day dark-matter (DM) abundance has garnered widespread attention. (See Refs.~\cite{Khlopov:2008qy,Carr:2020xqk,Green:2020jor,Carr:2020gox,Villanueva-Domingo:2021spv,Escriva:2022duf,Carr:2023tpt} for recent reviews.) Cosmologists have identified various mechanisms to amplify primordial overdensities that could undergo gravitational collapse into PBHs, to yield an appropriate population that addresses the DM mystery \cite{Belotsky:2014kca,Carr:2020gox,Escriva:2022duf,Ozsoy:2023ryl,Carr:2023tpt}. 

The PBH mass at the peak of the mass distribution, $\bar{M} (t_c)$, is proportional to the mass enclosed within a Hubble volume $M_H (t_c)$ at the time of the PBHs' formation, $t_c$ \cite{Carr:1975qj,Niemeyer:1997mt,Green:1999xm,Kuhnel:2015vtw}. This establishes a relationship between the typical PBH mass and the time of collapse. A combination of theoretical and observational bounds leaves a window $10^{17} \, {\rm g} \leq \bar{M} (t_c) \leq 10^{22} \, {\rm g}$ within which PBHs could constitute all of dark matter today \cite{Carr:2020xqk,Green:2020jor,Carr:2020gox,Villanueva-Domingo:2021spv,Escriva:2022duf}, which in turn constrains the time of collapse to $10^{-21} \, {\rm s} \leq t_c \leq 10^{-16} \, {\rm s}$. 

At these very early times, the plasma filling the universe had a temperature $10^5 \, {\rm GeV} \leq T (t_c) \leq 10^7 \, {\rm GeV}$, exponentially greater than the QCD confinement scale $\Lambda_{\rm QCD} = 0.17 \, {\rm GeV}$. At such high temperatures, the quarks and gluons in the plasma were {\it unconfined} \cite{Mukhanov:2005sc,Boyanovsky:2006bf,Kapusta:2006pm}. Therefore PBHs of relevance to DM necessarily formed by absorbing large collections of quarks and gluons from the quark-gluon plasma (QGP), which were not confined within color-neutral hadronic states.

Non-Abelian dynamics among the unconfined quarks and gluons yield a nontrivial distribution of QCD color charge within the QGP \cite{Blaizot:2001nr,Litim:2001db,Mrowczynski:2016etf}. In particular, collective modes of soft gluons, with momenta $k_{\rm soft} \sim g_s T$, where $g_s$ is the dimensionless gauge coupling strength, can produce spatial regions of nonvanishing net color charge, whose typical size is set by the Debye screening length $\lambda_D (T) \sim 1 / (g_s T)$ \cite{Manuel:2003zr,Manuel:2004gk}.

The mechanism by which PBHs form, known as ``critical collapse," indicates that some PBHs would form with arbitrarily small masses, $M \ll \bar{M}$ \cite{Choptuik:1992jv,Evans:1994pj,Niemeyer:1999ak,Gundlach:2002sx,Gundlach:2007gc,Musco:2008hv,Escriva:2021aeh}. The long-lived PBHs with mass $\bar{M}$ would constitute DM today, whereas those with $M\ll\bar{M}$ would have already evaporated. Nonetheless, these small PBHs would have formed from collapse regions of size $\sim\lambda_D$, on scales for which the QGP can have a nontrivial color-charge distribution. Such PBHs would form with net QCD color charge; a subset of these, moreover, would form with extremal enclosed charge, ${\cal Q} = \sqrt{G} \, M$. On the other hand, DM candidates (with mass $\sim\bar{M}$) would be color neutral.

Although vacuum solutions of color-charged black holes have been known in the literature for some time \cite{Volkov:1998cc,Volkov:2016ehx}, such simple scenarios have ignored how these objects could have formed. In contrast, we focus on realistic, well-motivated mechanisms by which a population of charged black holes---including near-extremal ones---could have formed amid a nontrivial medium in our actual universe.

The abundance of near-extremal PBHs depends sensitively upon the ratio $M/ \bar{M}$, which scales with the temperature of the plasma at the time of PBH formation, and hence falls over time as the plasma cools. Although rare and short-lived on cosmological timescales, these near-extremal charged PBHs would constitute an entirely new state of matter, with enclosed QCD charge ${\cal O} (10^{13} \, g_s)$, unlike the multi-parton states following relativistic heavy ion collisions, which briefly involve ${\cal O} (10^2)$ unconfined charges \cite{Pasechnik:2016wkt}. Moreover, as discussed below, these unusual PBHs could have phenomenological implications \cite{UnitsNote}.

{\it PBHs and Critical Collapse.} We restrict attention to PBHs that form from collapse of primordial overdensities amplified during inflation, which remains the most thoroughly analyzed and empirically constrained scenario \cite{Carr:2020gox,Escriva:2022duf,Ozsoy:2023ryl,Green:2020jor,Carr:2020xqk,Khlopov:2008qy,Villanueva-Domingo:2021spv,Belotsky:2014kca}.
One of the most striking results from decades of studies by the numerical-relativity community is that the mass of the resulting black hole depends upon a one-parameter family of initial data and a single universal critical exponent, in close analogy to phase transitions in statistical physics \cite{Choptuik:1992jv,Evans:1994pj,Niemeyer:1999ak,Musco:2008hv,Gundlach:2002sx,Gundlach:2007gc}. In particular, the black hole mass $M$ at the time of collapse $t_c$ obeys the relation
\beq
\label{MBH}
M (t_c)=\kappa M_H(t_c) \lvert \bar{\cal C} - {\cal C}_c\rvert^\nu,
\eeq
where $M_H(t_c)$ is the mass contained within a Hubble volume at $t_c$, ${\cal C}(r)=2G\delta M(r)/r$ is the compaction as a function of areal radius $r$ \cite{Shibata:1999zs,Harada:2023ffo}, $\bar{\cal C}$ is the compaction averaged over a Hubble radius, ${\cal C}_c \simeq 0.4$ is the threshold for black hole formation \cite{Escriva:2019phb}, and $\kappa$ is an ${\cal O}(1)$ dimensionless constant whose value depends on the spatial profile of ${\cal C}(r)$ and the averaging procedure \cite{Musco:2008hv,Musco:2018rwt,Ando:2018qdb,Germani:2018jgr,Kalaja:2019uju,Escriva:2019nsa,Young:2019osy,Gow:2020bzo,Escriva:2022duf}. The universal scaling exponent $\nu$ depends on the equation of state of the fluid that undergoes collapse; numerical studies show $\nu = 0.36$ for a radiation fluid \cite{Evans:1994pj,Koike:1995jm,Niemeyer:1999ak,Musco:2008hv}. Early numerical studies only considered spherically symmetric initial conditions, though later studies identified critical-collapse behavior even when relaxing spherical symmetry \cite{Gundlach:2007gc}. 

Researchers have distinguished ``Type I'' versus ``Type II'' forms of critical collapse. In Type I cases, a mass gap appears, setting a smallest (nonzero) value of $M$, above which $M$ scales as in Eq.~(\ref{MBH}). In Type II cases, the system remains scale-free and self-similar, with no mass gap, and self-consistent black hole solutions exist even for $M\rightarrow 0$ \cite{Gundlach:2002sx,Gundlach:2007gc}. Simulations have shown that several cases of cosmological interest are Type II, including collapse within a perfect fluid in the ultrarelativistic regime, as well as in a pure SU($N$) gauge field \cite{Choptuik:1996yg,Neilsen:1998qc,Choptuik:1999gh,Musco:2008hv}. 

Critical collapse yields a mass distribution $\psi(M, t_c)\equiv \rho_{\rm PBH}^{-1}(t_c)\,\dv*{\,n_{\rm PBH}(M,t_c)}{\,\rm{ln} \,M}$ that is strongly peaked near
\beq
\label{Mbar}
\bar{M}(t_c)=\gamma \,M_H(t_c)
\eeq
with $\gamma\simeq 0.2$. Given the probability distribution function $P (\bar{\cal C})$ for 
$\bar{\cal C}$, as we will see below, the mass distribution $\psi (M, t_c)$ features 
a power-law tail for masses $M\ll \bar{M}$ \cite{Carr:1975qj,Niemeyer:1997mt,Green:1999xm,Kuhnel:2015vtw}. Thus, for Type II collapse, some PBHs will form with arbitrarily small $M$ \cite{TypeIIb}. The radii of the PBHs of interest are so small that Bondi accretion remains negligible over relevant time-scales \cite{Rice:2017avg,DeLuca:2020fpg,Alonso-Monsalve:2023jfq} (see Supplemental Materials \cite{PRLSM}).

Another consequence of the scale-free, self-similar behavior of Type II collapse is that for initial data near the critical value, $\bar{\cal C} \simeq {\cal C}_c$, any dimensionful quantity related to the resulting black hole must scale with the same universal exponent $\nu$ as the mass \cite{Gundlach:2002sx,Gundlach:2007gc}. For such solutions, the initial stages of collapse begin with a region of radius $r \sim 1 / H$, but as time evolves, there is a net outflow of matter. The critical scaling ensures that the mass enclosed in the collapse region $r(t)$ scales as $M_{\rm enc} (r(t)) / r(t) = {\rm constant}$, as observed in numerical simulations \cite{Evans:1994pj,Niemeyer:1997mt}. (Note that this is different from estimates outside the context of critical collapse, for which one would expect $M_{\rm enc} \sim r^3$.) This is why black holes that result from an initial compaction $\bar{\cal C} \simeq {\cal C}_c$ have masses $M \ll \bar{M} \sim M_H$. At the time of collapse $t_c$, black hole formation is triggered and the mass outflow ceases; we call the final collapse radius $r_c \equiv r (t_c)$. Physically, all the matter contained within the volume of radius $r_c$ becomes contained within the black hole, $M_{\rm enc} (r_c) = M$.

Given the bound $10^{17} \, {\rm g} \leq \bar{M} (t_c) \leq 10^{22} \, {\rm g}$ within which PBHs could constitute all of dark matter \cite{Green:2020jor,Carr:2020gox,Carr:2020xqk,Villanueva-Domingo:2021spv,Escriva:2022duf}, and the relationship in Eq.~(\ref{Mbar}), the collapse times are constrained to $10^{-21} \, {\rm s} \leq t_c \leq 10^{-16} \, {\rm s}$. These times are exponentially earlier than the QCD confinement transition, which occurred at $t_{\rm QCD}\simeq 10^{-5}\,{\rm s}$, when the temperature of the plasma $T$ became comparable to the QCD scale $\Lambda_{\rm QCD}=0.17\,{\rm GeV}$. As noted above, if PBHs constitute a significant fraction of all DM, they must have formed amid a hot QGP.

{\it QGP and Debye Screening.} At temperatures $T\gg \Lambda_{\rm QCD}$, the fluid filling the universe was a hot plasma dominated by unconfined quarks and gluons. Within this hot QGP, color-charged particles such as soft gluons with momenta $k_{\rm soft}\sim g_s T$ undergo (non-Abelian) Debye screening, with screening length $\lambda_D(T)\sim 1/(g_s T)$. (For reviews, see Refs.~\cite{Pisarski:1988vd,Blaizot:2001nr,Litim:2001db,Kapusta:2006pm,Bazavov:2020teh,Mrowczynski:2016etf}.) This holds even within an expanding Friedmann-Lema\^itre-Robertson-Walker (FLRW) spacetime, given the large hierarchy of scales $\lambda_D\ll H^{-1}$. (For example, at $t_c = 10^{-21} \, {\rm s}$, when the plasma temperature was $T \simeq 10^7 \, {\rm GeV}$, one has $\lambda_D \sim 10^{-23} \, {\rm m}$ and $H^{-1} \sim 10^{-12} \, {\rm m}$ \cite{Alonso-Monsalve:2023jfq}.) This implies that at some distance $r$ from a charged particle, the effective charge falls as $\exp[-r/\lambda_D]$.

To quantify these effects, we construct an effective field theory for soft gluon modes, whose occupation numbers dominate the plasma. We decompose the full Yang-Mills gauge field $\bar{A}^c_\mu$ into hard ($a^c_\mu$) and soft ($A^c_\mu$) modes, with typical momenta $k_{\rm hard} \sim T$ and $k_{\rm soft} \sim g_s T$, respectively, where $c=1,\ldots,8$ is the color index. The fields $A^c_\mu$ have large occupation numbers per mode, so they behave as effectively classical fields. The dynamics of these soft modes are governed on length-scales $\lambda\geq 1/k_{\rm soft}$ by an effective action \cite{Alonso-Monsalve:2023jfq}
\beq
S_{\rm eff} = \int d^4 x \sqrt{-g} \left[ \frac{ M_{\rm pl}^2}{2} R - \frac{1}{4} F_{\mu\nu}^a F_a^{\mu\nu} - j_\mu^a A_a^\mu  + {\cal L}_{\rm fluid}\right] ,
\label{Seff}
\eeq
which results from integrating out the hard modes. We follow the background-field method, which allows us to choose the gauge symmetry to act only on $a_\mu^c$. As described in detail in Ref.~\cite{Blaizot:2001nr} (see also Ref.~\cite{Alonso-Monsalve:2023jfq}), one may then choose a gauge (for example, by including a gauge-fixing term such as $\nabla_i a^i_a - g_s f^{abc} A_i^b a_c^i$ in the full theory) such that no additional gauge-fixing terms or ghosts are required in $S_{\rm eff}$. Here $F_{\mu\nu}^a=\nabla_\mu A^a_\nu-\nabla_\nu A^a_\mu + g_s f^{abc}A^b_\mu A^c_\nu$ is the field strength for soft modes, with $\nabla_\mu$ the spacetime covariant derivative and $f^{abc}$ the SU(3) structure constants; $j_\mu^a (x)$ is the current induced by (non-Abelian) self-interactions with the hard modes (including high-momentum quarks); and ${\cal L}_{\rm fluid}$ represents contributions to the spacetime evolution from constituents other than the soft modes.

The induced current $j^\mu_a (x)$ depends on the deviation from an equilibrium distribution function for the high-energy charged particles \cite{Blaizot:2001nr,Litim:2001db,Alonso-Monsalve:2023jfq}. For spacetimes for which one can set $g_{0i}=0$, and in the approximately static limit $\lvert \partial_0 g^{\mu\nu}\rvert,\, \lvert \partial_0 A_\mu^a\rvert / \lvert A^a_\mu\rvert \ll k_{\rm soft}$, one finds \cite{Alonso-Monsalve:2023jfq}
\beq
j^a_\mu = m_D^2\,A_0^a\,\delta^0_\mu + {\cal O}(g_s^3),
\label{j}
\eeq
in terms of the Debye mass $m^2_D = (2N_c+N_f)g_s^2 T^2 / 6 \sim g_s^2 T^2$, where $N_c = 3$ is the number of colors, $N_f$ is the number of effectively massless quarks, and $T$ is the Tolman temperature \cite{Alonso-Monsalve:2023jfq}. The Debye length is given by $\lambda_D \equiv 1/m_D$.

Comparing Eqs.~(\ref{Seff}) and (\ref{j}) reveals that the chromoelectric component $A_0^a$ acquires an effective mass, while the chromomagnetic components $A_i^a$ remain massless. The equations of motion that follow from $S_{\rm eff}$ are $\nabla^\mu F^a_{\mu\nu} + g_s f^{abc}A_b^\mu F^c_{\mu\nu} = m_D^2 \,A_0^a \,\delta^0_\nu$. In an FLRW background, a closed-form analytic solution for the components $A_\mu^a (x)$ for a point charge may be found \cite{Alonso-Monsalve:2023jfq}. We may further identify a quasi-local chromoelectric charge based on the form of the gauge-invariant quantity $g^{ij}\delta_{ab} E_i^a E_j^b = {\cal Q}^2 (r,T) / r^4$, where $E_i^a = F_{0i}^a$ is the chromoelectric field. This yields \cite{Alonso-Monsalve:2023jfq}
\beq
{\cal Q} (r, T) = {\cal Q}_0 \left( 1 + \frac{r}{\lambda_D} \right) e^{- r / \lambda_D} ,
\label{Qenc}
\eeq
where the charge at the origin, ${\cal Q}_0 = [ \delta_{ab} {\cal Q}^a_0 {\cal Q}^b_0 ]^{1/2}$, is an integer multiple of the unit charge $g_s$, and $\lambda_D = \lambda_D (T)$.

Given Debye screening, the QGP will be color neutral on long length scales $r\gg\lambda_D$, but can have a nontrivial distribution of color charge across shorter length scales $r\sim \lambda_D$. In particular, there can exist regions with net color charge, whose spatial extent is set by $\lambda_D(T)$ \cite{Manuel:2003zr,Manuel:2004gk}. We may estimate the color charge inside one such region in terms of the number of soft gluons enclosed, with each contributing a unit charge $g_s$ in the same direction in color space. These soft gluons will represent a fraction $F$ of all the particles contained within a particular net-color region (including hard gluons and other particle species), given by
\beq
F\equiv \frac{n^{\text{cc}}_{\rm soft}}{n_{\rm total}}\simeq \frac{3\pi}{2}\frac{\alpha_s}{g_\ast}\sim {\cal O}(10^{-3}),
\label{Fa}
\eeq
where $\alpha_s \equiv g_s^2/(4\pi)$ and $g_\ast$ is the number of effectively massless degrees of freedom; at the energy scales of interest, $T> m_t = 173\,{\rm GeV}$, $g_\ast=106.75$ for the Standard Model. Eq.~(\ref{Fa}) follows from evaluating the number density $n_{\rm soft}^{\rm cc}$ of soft gluons within a single spatial region of size $\lambda_D$
by truncating the momentum integral over the distribution function at $k_{\rm soft}$.

The total color charge in a spatial volume depends on the number of distinct net-color regions enclosed. As a conservative estimate, we assume that only one such region located at the center contributes to ${\cal Q}_0^a$ inside a spherical volume of plasma of radius $r_c\geq \lambda_D$, while the rest of the plasma in the volume screens ${\cal Q}_0^a$. Moreover, we approximate ${\cal Q}_0^a$ as a point charge, which again underestimates the net charge by overestimating the effect of screening. We may exploit the gauge symmetry to assign a specific color (e.g., $a = 1$) to the soft gluons within the central net-charge region; within that central region, we approximate the number density for soft gluons of all other charges ($a = 2, \dots , 8$) to vanish. Then the net color charge ${\cal Q}(r_c, T)$ in a volume of plasma of radius $r_c$ may be approximated as
\beq
{\cal Q} (r_c,T)\simeq g_s\, F\, N_{\text{cc}}\left( 1 + \frac{r_c}{\lambda_D} \right) e^{-r_c/\lambda_D},
\label{Qnet}
\eeq
where $N_{\text{cc}}$ is the total number of particles in the central region of net color charge.

{\it PBHs with Significant QCD Charge.} In order to quantify the net enclosed charge ${\cal Q}(r_c, T_c)$, we estimate $N_{\text{cc}}$ for a PBH that forms with mass $M$ at temperature $T_c$ as
\beq
    N_{\text{cc}} \simeq \frac{\lambda_D^3}{r_c^3}\frac{M (r_c, T_c)}{T_c},
\label{Ncc}
\eeq
using the fact that in a radiation bath in thermal equilibrium the average energy per particle is $\sim T_c$, and the number of distinct net-color regions absorbed by the PBH scales as $(r_c / \lambda_D)^3$ \cite{TempNote}. From Eqs.~(\ref{Fa})--(\ref{Ncc}), the net QCD charge contained within a PBH of mass $M$ that forms at time $t_c$ is then
\beq
{\cal Q}_{\rm PBH} (r_c, T_c) \simeq \frac{3 g_s^3}{8 g_*} \left( \frac{ 1 + {\cal R}_c}{{\cal R}_c^3} \right) e^{- {\cal R}_c} \frac{M (r_c, T_c)}{T_c} ,
\label{QPBH}
\eeq
where we have defined the dimensionless ratio ${\cal R}_c \equiv r_c / \lambda_D$. The net enclosed charge grows inversely with $T_c$, since the average energy per particle falls with $T_c$, so a larger number of charge-carrying particles must be absorbed to form a PBH of mass $M$ at lower temperatures. On the other hand, as expected, the net charge falls rapidly for ${\cal R}_c \gg 1$, given screening within the medium prior to PBH collapse.

Based on exact black hole solutions of the Einstein-Yang-Mills equations in vacuum, which are analogous to Reissner-Nordstr\"om black holes with electromagnetic charge, an extremal color-charged PBH satisfies ${\cal Q}_{\rm extr} = \sqrt{G} \, M_{\rm extr}$ \cite{Volkov:1998cc,Volkov:2016ehx}. For simplicity, we adopt this value for the maximal charge ${\cal Q}_{\rm extr}$ even in the presence of plasma external to the PBH. Using Eq.~(\ref{QPBH}), we then find that an extremal PBH forms with (dimensionless) collapse radius ${\cal R}_{\rm extr} \equiv r_c^{\rm extr} / \lambda_D$ given by
\beq
\left( \frac{1 + {\cal R}_{\rm extr}}{{\cal R}_{\rm extr}^3} \right) e^{- {\cal R}_{\rm extr}} = \sqrt{ \frac{ 8 \pi}{9}} \frac{ g_*}{g_s^3} \frac{ T_c}{M_{\rm pl}} .
\label{Rextr}
\eeq
The radius ${\cal R}_{\rm extr}$ is uniquely determined by the temperature of the plasma at the time of collapse, $T_c$.

As noted above, the collapse radius $r_c$ and the mass $M$ of a resulting PBH scale similarly, $r_c \sim G M \sim \lvert \bar{\cal C} - {\cal C}_c \rvert^\nu$, due to critical collapse \cite{Evans:1994pj,Niemeyer:1997mt}. Recent numerical simulations have found that the scaling of Eq.~(\ref{MBH}) remains accurate (deviating by less than $15\%$) even near the peak of the mass distribution, for initial conditions far from the critical value
\cite{Escriva:2019nsa}. We may therefore compare $r_c$ and $M$ for small PBHs with those at the peak of the mass distribution, $\bar{r}_c$ and $\bar{M}$. Given Eq.~(\ref{Mbar}), we find $\bar{r}_c \simeq 1 / (2 H (t_c))$, consistent with Carr's original analytic estimate \cite{Carr:1975qj}. For $M < \bar{M}$, the critical scaling $r_c \sim G M$ then yields
\beq
\frac{ M (r_c, T_c)}{\bar{M} (T_c)} \simeq \frac{ 2 \pi}{g_s} \sqrt{ \frac{ g_*}{90}} \frac{ T_c}{ M_{\rm pl}} {\cal R}_c ,
\label{MtobarM}
\eeq
upon making use of $\lambda_D \simeq 1 / (g_s T_c)$ and $H^2 (t_c) = (\pi^2 g_* / 90) \, T^4_c / M_{\rm pl}^2$ to relate the Hubble parameter to the fluid temperature at the time of collapse. PBHs that form with ${\cal R}_c \sim {\cal O} (1)$ will be closer to the peak-mass $\bar{M}$ at higher temperatures (and hence at earlier times) than those that form at lower temperatures. 

We may combine Eqs.~(\ref{Rextr}) and (\ref{MtobarM}) to estimate the abundance of near-extremal PBHs as a function of the collapse time $t_c$. As described in the Supplemental Materials \cite{PRLSM}, the mass distribution then becomes
\beq
    \psi(M, t_c) = \left( \frac{1}{\rho_{\rm PBH} {V}(t_c)} \right)\frac{m^{1/\nu}}{\nu} P({\cal C}_c+m^{1/\nu}),
\eeq
where $m\equiv M / [\kappa M_H(t_c)]$, and $P (\bar{\cal C})$ is the probability distribution function for the (spatially averaged) compaction $\bar{\cal C}$. As noted above, PBHs with near-extremal charge will form with $M \ll \bar{M}$, or $m \ll 1$. In that limit, we may integrate $\dv*{n_{\rm PBH}(M,t_c)}{M} = \rho_{\rm PBH} \, \psi (M, t_c) / M$ to obtain the number of PBHs per Hubble volume at time $t_c$ with masses up to some value $\hat{M} \ll \bar{M}$. This yields
\beq
    N_{\rm PBH}(\text{up to }\hat{M},t_c)\simeq \sqrt{\frac{2}{\pi\sigma^2}}\exp[-\frac{{\cal C}_c^2}{2\sigma^2}]\left( \frac{\gamma}{\kappa} \frac{\hat{M}}{\bar{M}(t_c)} \right)^{1/\nu}.
    \label{Npbh}
\eeq
As discussed in the Supplemental Material \cite{PRLSM}, non-Gaussian features modify $P ( \bar{\cal C})$ near the peak $\bar{M}$ but remain negligible in the limit $m \ll 1$. In that limit, $P (\bar{\cal C})$ is well approximated by the usual Gaussian Press-Schechter form, with variance $\sigma^2$ in the range of interest $10^{-2} \leq \sigma^2 \leq 10^{-1}$ \cite{PRLSM}. By combining Eqs.~(\ref{Rextr}) and (\ref{MtobarM}), we may evaluate Eq.~(\ref{Npbh}) for the case of extremal PBHs, for $\hat{M} \rightarrow M_{\rm extr}$.

\begin{figure}
    \centering
    \includegraphics[width=3.2in]{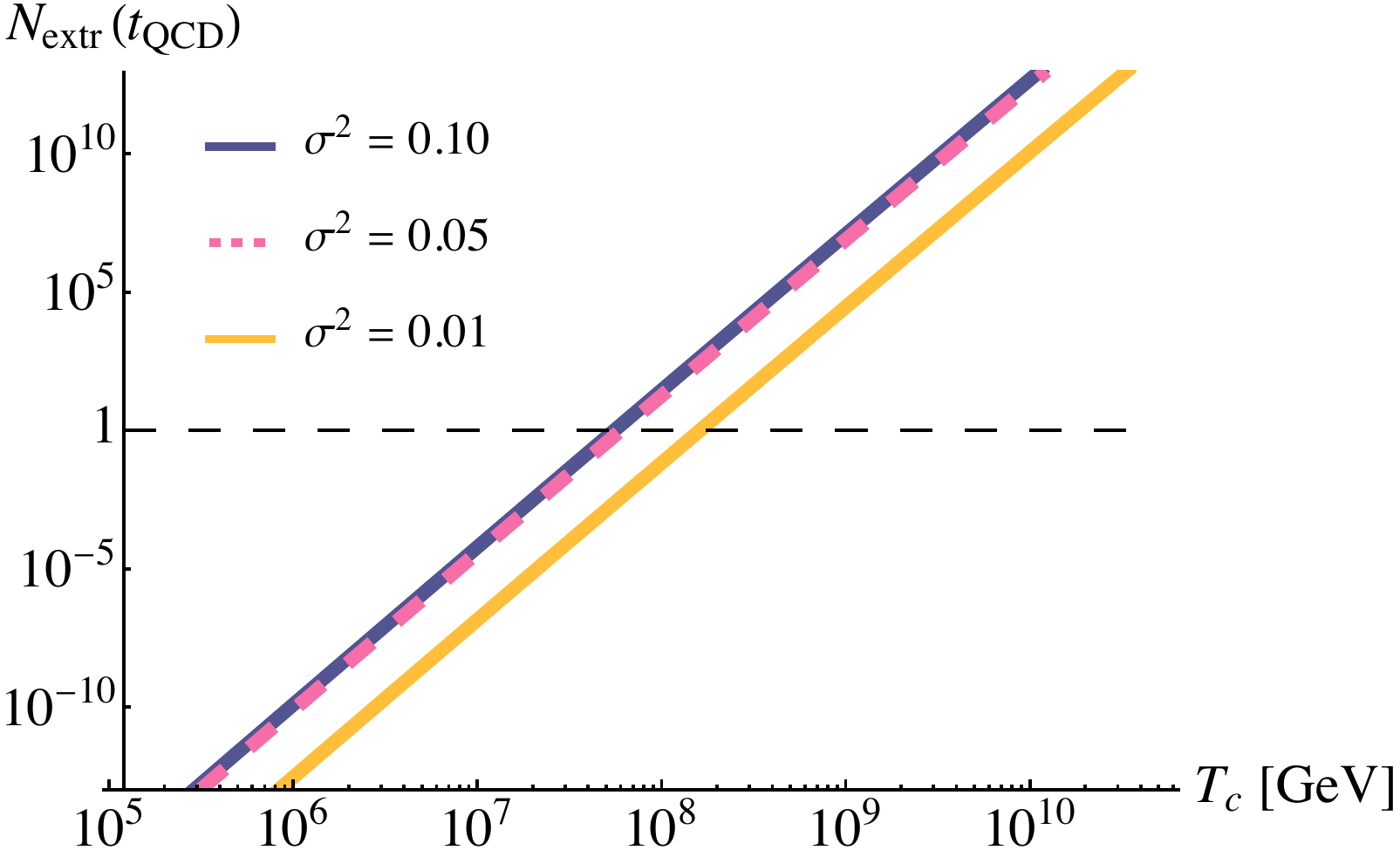}
    \caption{The number $N_{\rm extr}$ of PBHs with extremal QCD color charge per Hubble volume at the time of the QCD confinement transition, $t_{\rm QCD} = 10^{-5} \, {\rm s}$, as a function of the plasma temperature $T_c$ at the time the PBHs form, for different values of the variance of the spatially averaged compaction, $10^{-2} \leq \sigma^2 \leq 10^{-1}$. We have set $\kappa = 1$, $\gamma = 0.2$, ${\cal C}_c = 0.4$, and $\nu = 0.36$.}
    \label{fig:Nextr}
\end{figure}

At later times $t>t_c$, the total number of PBHs in a Hubble sphere grows as $[a(t)/a(t_c)]^3$, due to the expansion of the universe. Since $a(t)\propto t^{1/2}$ during radiation-dominated expansion, the total number of PBHs within a Hubble sphere at time $t$ is $N_{\rm PBH}(\text{up to }\hat{M},t)=N_{\rm PBH}(\text{up to }\hat{M},t_c)(t/t_c)^{3/2}$. As shown in Fig.~\ref{fig:Nextr}, we expect $N_{\rm extr} (t_{\rm QCD}) \simeq 1$ extremal PBHs per Hubble volume at the time of the QCD confinement transition, $t_{\rm QCD} = 10^{-5} \, \rm s$, for a population of PBHs that forms amid a plasma of temperature $T_c \simeq 5 \times 10^7 \, {\rm GeV}$.

The plasma temperature $T_c \simeq 5 \times 10^7 \, {\rm GeV}$ is remarkably close to the window within which PBHs could account for all of dark matter today, $10^{5} \, {\rm GeV} \leq T_c \leq 10^7 \,{\rm GeV}$. In particular, $\bar{M} (T_c = 5 \times 10^7 \, {\rm GeV}) = 7 \times 10^{15} \, \rm g$, for which the expected lifetime due to Hawking evaporation $t_{\rm evap} \simeq 10^{22} \, {\rm s}$ exponentially exceeds the age of our observable universe. Meanwhile, an extremal black hole that forms at $T_c = 5 \times 10^7 \, {\rm GeV}$ has a mass $M_{\rm extr} = 2 \times 10^7 \, {\rm g}$ and charge ${\cal Q} = M_{\rm extr} / M_{\rm pl} \sim {\cal O}(10^{13} \, g_s)$. The evaporation lifetime of an {\it uncharged} (Schwarzschild) black hole of the same mass is $t_{\rm evap} \simeq 10^{-4} \, {\rm s} > t_{\rm QCD}$. Given that the Hawking temperature $T_H \rightarrow 0$ as a black hole approaches extremality, we expect $t_{\rm evap} \gg t_{\rm QCD}$ for near-extremal PBHs that form amid a plasma at $T_c = 5 \times 10^7 \, {\rm GeV}$.

In general, a PBH should form with significant net color charge if $r_c \sim \lambda_D \sim 1 / (g_s T)$.
Because of the special scaling features of self-similar critical-collapse solutions, we further expect $M_{\rm extr} \sim r_c$; whereas the peak-mass $\bar{M} \sim 1 / H(t_c) \sim 1 / T_c^2$. Hence the ratio $M_{\rm extr} / \bar{M} \sim T_c$ falls over time, as the plasma cools. The number of extremal PBHs scales as $(M_{\rm extr} / \bar{M} )^{1/\nu}$, as per Eq.~(\ref{Npbh}). For $T_c \ll M_{\rm pl}$, this is a small number raised to a large power, confirming the pattern shown in Fig.~\ref{fig:Nextr}.

Finally we note that our effective field theory for soft gluons remains self-consistent for $\lambda_D (T) \leq r_c^{\rm extr} (T) \ll H^{-1} (T)$. For a hot plasma filled with Standard Model particles, $\lambda_D (T) \simeq 10^{-31} \, {\rm m} \, ( 10^{15} {\rm GeV} / T )$ and $H^{-1} (T) \simeq 10^{-29} \, {\rm m} \,(10^{15} \, {\rm GeV} / T )^2 $, indicating that $\lambda_D \ll H^{-1}$ for all $T \leq 10^{15} \, {\rm GeV}$, i.e., below the limit on the post-inflation reheating temperature set by the present bound on the primordial tensor-to-scalar ratio 
\cite{BICEP:2021xfz}. In addition, Eq.~(\ref{Rextr}) confirms that $\lambda_D \leq r_c^{\rm extr}$ for all $T \leq 10^{16} \, {\rm GeV}$.

{\it Discussion.} Primordial black holes remain a tantalizing candidate with which to address the long-standing mystery of dark matter. Present-day constraints on such a scenario require that the relevant population of PBHs be produced at very early times, exponentially earlier than the time of the QCD confinement transition at $t_{\rm QCD} = 10^{-5}$ s. Combining the well-studied phenomenon of critical collapse---which can produce PBHs of arbitrarily small masses---with the nontrivial charge distribution of soft gluons within a hot quark-gluon plasma suggests that some PBHs with net QCD color charge are likely to form at times $t_c < t_{\rm QCD}$. The total number of such charged PBHs per Hubble volume depends sensitively on details like their time of formation.

The evolution of a near-extremal PBH after its formation comprises two periods: net absorption (of mass and compensating charges) starting at $t_c$, followed by Hawking emission. The change occurs at a time $t_{\rm equal}$ when the temperature of the surrounding plasma matches the Hawking temperature of the PBH, as the former decreases with time and the latter increases due to the absorption of compensating charges. (The increase in mass accumulated during $\Delta t= t_{\rm equal}-t_c$ remains an exponentially small fraction of the original PBH mass, and therefore does not lower the Hawking temperature appreciably \cite{Rice:2017avg,DeLuca:2020fpg,Alonso-Monsalve:2023jfq,PRLSM}.) 

Neither process is likely to completely discharge a near-extremal PBH. First, the PBH's charge is strongly screened in the medium \cite{Alonso-Monsalve:2023jfq}, reducing the probability of the PBH preferentially absorbing compensating charges. Second, local temperature gradients in the plasma near the PBH \cite{Alonso-Monsalve:2023jfq} reduce the number of 
particles absorbed for a given accreted mass $\Delta M$. (See the Supplemental Materials \cite{PRLSM} for details on the exponentially inefficient accretion for the PBHs of interest.) 
We therefore expect a lower bound on the discharge time-scale to be set by the Hawking evaporation time, which, as noted above, satisfies $t_{\rm evap} > t_{\rm QCD}$ for the PBHs of interest. An additional complication that requires further research is how a PBH that is charged under both the fundamental and adjoint representations of SU(3) could discharge following $t_{\rm QCD}$.

Assuming that charged PBHs immersed within the QGP remain stable at least until $t_{\rm QCD} = 10^{-5} \, {\rm s}$, the question of what happens following $t_{\rm QCD}$ becomes critical \cite{StabilityNote}. After the QCD confinement transition, the medium ceases to screen the PBH's enclosed charge ($\lambda_D \rightarrow \infty$), and it becomes energetically (very) costly for the PBH to maintain its charge.

A likely scenario following $t_{\rm QCD}$ is that the gravitational potential energy of the PBH would induce a cloud of color-charged (virtual) particles from the vacuum, forming a gray-body penumbra of radius $R_{\rm gb}$. We estimate $R_{\rm gb}$ by comparing the local particle acceleration to the QCD scale, $G M / R_{\rm gb}^2 \sim \Lambda_{\rm QCD}$, or
\beq
R_{\rm gb} \sim \sqrt{ r_{\rm PBH} \cdot \ell_{\rm QCD}} ,
\label{Rgb}
\eeq
where $r_{\rm PBH} \sim G M$ is the radius of the outer trapping surface of the black hole and $\ell_{\rm QCD} \sim 10^{-15} \, {\rm m}$ is the length-scale associated with $\Lambda_{\rm QCD} = 0.17 \, {\rm GeV}$. Thereafter, the PBH and its cloud would behave as a color-neutral ``hadron." (See also Ref.~\cite{Feng:2022evy}.) For PBHs that form near the window $10^{-21} \, {\rm s} \leq t_c \leq 10^{-16} \, {\rm s}$ relevant for dark matter, such hadrons would have radii about four orders of magnitude smaller than the radius of a proton, but masses more than 30 orders of magnitude greater than the mass of a proton.

The rapid transition around $t_{\rm QCD}$ might produce gravitational waves. The ``dressing" of a PBH with a charge-canceling cloud could induce local density gradients that would source scalar metric perturbations, which in turn would induce tensor perturbations at second order \cite{Domenech:2021ztg}; the frequency of such a signal would likely peak near $f \sim 1 / R_{\rm gb} \sim 10^{27} \, {\rm Hz}$, well beyond sensitivities of projected detectors. However, if some of these objects persisted well beyond $t_{\rm QCD}$, they might yield observable effects. For example, it is possible that they could disrupt the thermal equilibrium distribution of protons and neutrons around the onset of big-bang nucleosynthesis. (See also Refs.~\cite{Kawasaki:2004qu,Carr:2009jm,deFreitasPacheco:2023hpb}.) For extremal PBHs that form amid a plasma at $T_c \sim 5 \times 10^7 \, {\rm GeV}$, there would exist $\sim 10^7$ extremal PBHs (plus clouds) per Hubble volume at the onset of nucleosynthesis. Any such signature would enable a first-of-its-kind probe of the small-mass tail of the PBH distribution.

Beyond these potentially observable consequences, the scenario described here could have implications for the no-hair theorem and cosmic censorship. Whether the presence of a screening medium impacts the non-Abelian hair previously found for QCD-charged black holes in vacuum \cite{Coleman:1991sj,Coleman:1991jf,
Coleman:1991ku,Coleman:1992rn,Krauss:1996df,GarciaGarcia:2018tua,Volkov:1998cc,Volkov:2016ehx} requires further study. Other studies, meanwhile, have identified situations---such as charged black holes in an Einstein-Maxwell holographic model embedded in asymptotically anti-de Sitter spacetime---in which the black holes can evolve into naked singularities as the temperature of the state is decreased to $T = 0$ \cite{Horowitz:2016ezu,Crisford:2017gsb,Horowitz:2019eum,Engelhardt:2019btp}. In our universe, Debye screening becomes less effective as the temperature of the QGP cools, increasing the PBH's effective charge seen by particles in the plasma. Whether these changes in the medium could push a QCD-charged PBH to become post-extremal remains the subject of further research.

\vskip 20pt

{\it Acknowledgements.} It is a pleasure to thank Iv\'an Agull\'o, Chris Akers, Peter Arnold, Josu Aurrekoetxea, Thomas Baumgarte, Jolyon Bloomfield, Nathaniel Craig, \r{A}smund Folkestad, Rikab Gambhir, Carsten Gundlach, Alan Guth, Daniel Harlow, Sergio Hern\'andez-Cuenca, Gary Horowitz, Scott Hughes, Edmond Iancu, Mikhail Ivanov, Patrick Jefferson, Jamie Karthein, Joshua Lin, Hong Liu, Cristina Manuel, J\'{e}r\^{o}me Martin, David Mateos, Govert Nijs, Krishna Rajagopal, Fernando Romero-L\'opez, Bruno Scheihing Hitschfeld, Phiala Shanahan, Jesse Thaler, Vincent Vennin, and Xiaojun Yao for helpful discussions. Portions of this work were conducted in MIT's Center for Theoretical Physics and supported in part by the U.~S.~Department of Energy under Contract No.~DE-SC0012567. EAM is also supported by a fellowship from the MIT Department of Physics.


%

\clearpage

\onecolumngrid
\appendix

\setcounter{equation}{0}
\setcounter{table}{0}
\setcounter{figure}{0}
\renewcommand{\theequation}{S\arabic{equation}}

\centerline{\large\bf Supplemental Material: }
\vspace{0.15cm}
\centerline{\large\bf Primordial Black Holes with QCD Color Charge}
\vspace{0.5cm}

\twocolumngrid

\section*{Relativistic Bondi Accretion}

When calculating accretion onto a black hole, one might consider effects of outward-directed radiation pressure, in this case arising from Hawking radiation. To be conservative, we neglect this effect and consider Bondi accretion, which only accounts for inflowing matter from the surrounding medium \cite{ShapiroTeukolsky1983}.

As noted in Ref.~\cite{Alonso-Monsalve:2023jfq}, Bondi accretion remains strongly suppressed for the primordial black holes (PBHs) considered here, consistent with the findings in Refs.~\cite{Rice:2017avg,DeLuca:2020fpg}. Bondi accretion describes the adiabatic flow of ambient fluid onto a spherically symmetric, stationary, nonrotating black hole of mass $M$. We follow the treatment of relativistic Bondi accretion in Refs.~\cite{ShapiroTeukolsky1983,Richards:2021zbr}.

We are interested in Bondi accretion around the time of PBH formation, amid a hot plasma with a radiation-like equation of state, $P = w \rho$ with $w = 1/3$, and hence a speed of sound in the fluid far from the PBH of $c_{s, \infty}^2 = w = 1/3$. The Bondi accretion rate is evaluated at a critical radius $r_B$ at which the fluid undergoes a transition from subsonic to supersonic flow, which is given by \cite{ShapiroTeukolsky1983,Richards:2021zbr}
\beq
r_B = \frac{ GM}{2} \frac{ \left( 1 + 3 c_s^2 (r_B) \right)}{c_s^2 (r_B)} ,
\label{rB}
\eeq
where $c_s (r_B)$ is the sound speed of the fluid at the critical radius. One may then integrate the relativistic Euler equation to find a relationship between $c_s (r_B)$ and $c_{s,\infty}$; for the equation of state $P = \rho / 3$, which yields \cite{ShapiroTeukolsky1983,Richards:2021zbr}
\beq
\left( 1 + 3 c_s^2 (r_B) \right) \left( 1 - 3 c_s^2 (r_B) \right)^2 = 0 ,
\label{csrB1}
\eeq
or $c_s^2 (r_B) = 1/3 = c_{s,\infty}^2$. Eq.~(\ref{rB}) then yields the Bondi radius
\beq
r_B = 3 G M ,
\label{rB2}
\eeq
ensuring that $r_B > r_s = 2 G M$.

The accretion rate comes from integrating the contintuity equation for the fluid, and is given by the surface area $(4 \pi r^2)$ times the energy density of the fluid $(\rho)$ times the radial component of the fluid 4-velocity $(u \equiv \vert u^r \vert )$, all evaluated at the Bondi radius \cite{ShapiroTeukolsky1983,Richards:2021zbr}:
\beq
\dot{M} = 4 \pi r_B^2 \, \rho (r_B) \, u (r_B) .
\label{dotM1}
\eeq
This may be written in terms of asymptotic quantities (for $r \rightarrow \infty$) as
\beq
\dot{M} = 4 \pi \lambda_{\rm GR} \left( \frac{ GM}{c_{s, \infty}^2}\right)^2 \rho_\infty \, c_{s, \infty} ,
\label{dotM2}
\eeq
where the relativistic accretion-rate eigenvalue is $\lambda_{\rm GR} = 1 / \sqrt{2}$ for a fluid with radiation equation of state \cite{Richards:2021zbr}. Upon using $c_{s, \infty}^2 = 1/3$ and $\rho_\infty = (\pi^2 / 30) g_* T_\infty^4$ for a radiation fluid, Eq.~(\ref{dotM2}) yields
\beq
\frac{ \dot{M} }{M^2} = \sqrt{\frac{3}{2}} \frac{\pi g_*}{160} \left( \frac{ T_\infty}{M_{\rm pl}} \right)^4  .
\label{dotM3}
\eeq
Note the scaling of the dimensionless quantity $\dot{M} / M^2 \sim (T_\infty / M_{\rm pl} )^4$. The PBHs of interest form amid a plasma with $T_\infty \sim {\cal O} (10^7) \, {\rm GeV}$, and hence $\dot{M} / M^2 \sim {\cal O} (10^{-44})$ soon after the PBHs form.

During the time period of interest, the temperature of the plasma (at spatial infinity) falls such that $T_\infty (t) \, a(t) = {\rm constant}$, where $a (t) = (t / t_c)^{1/2}$ is the scale factor during the radiation-dominated era, and $t_c$ is the time of PBH formation. We may then integrate Eq.~(\ref{dotM3}) to find the total mass $\Delta M$ accreted between $t_c$ and some later time $t_{\rm late}$ (taking, as a conservative estimate, $g_* \simeq {\rm constant}$):
\beq
\frac{\Delta M}{M_c} = \sqrt{\frac{3}{2}} \frac{\pi g_*}{160} \left( \frac{ T_{c, \infty}}{M_{\rm pl}^2} \right)^4 M_c t_c \left( 1 - \frac{ t_c}{t_{\rm late}} \right) ,
\label{DeltaM}
\eeq
where $M_c \equiv M (t_c)$ is the initial PBH mass. For the extremal PBHs we consider here, which form with $M_c = M_{\rm extr} = 2 \times 10^7 \, {\rm g}$ amid a hot plasma with (asymptotic) temperature $T_{c, \infty} = 5 \times 10^7 \, {\rm GeV}$ at collapse time $t_c = 10^{-21} \, {\rm s}$, Eq.~(\ref{DeltaM}) indicates that the total mass accretion at arbitrarily late times will be bounded by
\beq
\frac{\Delta M}{M_{\rm extr}} \leq 1.3 \times 10^{-9} .
\label{DeltaMextr}
\eeq

\section*{PBH Mass Distribution and Abundance}

We found above that $\lambda_D \sim 1 / T_c$ whereas $\bar{M} \sim 1 / H (t_c) \sim 1 / T_c^2$, where $\lambda_D$ is the Debye screening length in the plasma, $\bar{M}$ is the peak of the PBH mass distribution, and $T_c = T_\infty (t_c)$; hence we expect the ratio $M / \bar{M} \sim {\cal R}_c T_c$ to fall as the plasma temperature at the time of gravitational collapse, $T_c$, falls. (As in the main text, ${\cal R}_c \equiv r_c / \lambda_D$ is the dimensionless ratio of the collapse radius to the Debye screening length.) The abundance of PBHs with mass $M < \bar{M}$ depends sensitively upon the ratio $M / \bar{M}$.

The energy density of PBHs within a Hubble volume at the time of formation $t_c$ may be written
\beq
\rho_{\rm PBH} (t_c) = \int dM \, M \frac{ d n_{\rm PBH} (M, t_c)}{dM} ,
\label{rhoPBH1}
\eeq
where $n_{\rm PBH}$ is the number density of PBHs. Given Eq.~(1) in the main paper for the mass of a PBH that results from critical collapse, we may similarly write
\beq
\rho_{\rm PBH} (t_c) = \frac{1}{V (t_c)} \int d \bar{\cal C} \, M (\bar{\cal C} , t_c) \, P (\bar{\cal C}) ,
\label{rhoPBH2}
\eeq
where $V (t_c) = 4 \pi / (2 H^3 (t_c))$ is the volume of the Hubble sphere at time $t_c$, and $P (\bar{\cal C})$ is the probability distribution function for the (spatially averaged) compaction $\bar{\cal C}$. From Eq.~(1) in the main paper, we may write
\beq
\bar{\cal C} = {\cal C}_c + m^{1/\nu} ,
\label{calCm}
\eeq
where ${\cal C}_c \simeq 0.4$ is the critical threshold for collapse \cite{Escriva:2019phb} and 
\beq
m \equiv \frac{ M (\bar{\cal C}, t_c )}{\kappa \, M_H (t_c)} .
\label{mdef}
\eeq
Then Eq.~(\ref{rhoPBH2}) becomes
\beq
\rho_{\rm PBH} (t_c) = \frac{1}{V (t_c)} \int dM \, \frac{M \,  m^{\frac{1}{\nu} - 1}}{\nu \kappa \, M_H (t_c) } P ({\cal C}_c + m^{1/\nu} ) .
\label{rhoPBH3}
\eeq
Comparing Eqs.~(\ref{rhoPBH1}) and (\ref{rhoPBH3}), we may identify
\beq
\frac{ d n_{\rm PBH} (M, t_c )}{dM} = \frac{1}{ V (t_c)} \frac{ m^{\frac{1}{\nu} - 1}}{\nu \kappa M_H (t_c)} P ({\cal C}_c + m^{1/\nu} ) .
\label{dnPBHdM}
\eeq
The mass function $\psi (M, t_c)$ is defined as 
\beq
\psi (M, t_c) \equiv \frac{ M}{ \rho_{\rm PBH} (t_c)} \frac{ d n_{\rm PBH} (M, t_c)}{dM} .
\label{psinPBH}
\eeq
Upon using Eq.~(\ref{dnPBHdM}), we then find
\beq
    \psi(M, t_c) = \left( \frac{1}{\rho_{\rm PBH} {V}(t_c)} \right)\frac{m^{1/\nu}}{\nu} P({\cal C}_c+m^{1/\nu}).
\eeq
Throughout this discussion, we have made no assumptions about the form of $P (\bar{\cal C})$.

Non-Gaussian probability distribution functions $P (\bar{\cal C})$ modify $\psi(M, t_c)$ near the peak of the distribution $\bar{M}$ compared to predictions from the Gaussian Press-Schechter formalism \cite{Young:2019yug,DeLuca:2019qsy,Gow:2020bzo,DeLuca:2020ioi,Musco:2020jjb,Gow:2020cou,Biagetti:2021eep,Ferrante:2022mui,Gow:2022jfb,DeLuca:2022rfz,DeLuca:2023tun,Escriva:2022duf}, but yield the same long (power-law) tail in $\psi (M, t_c)$ for $M\ll \bar{M}$. (See, e.g., the figures in Ref.~\cite{Gow:2022jfb}.) This behavior is straightforward to understand upon Taylor expanding a given ansatz for $P (\bar{\cal C}) = P ({\cal C}_c + m^{1/\nu})$ in the limit $m \ll 1$. Since $m \leq 10^{-9}$ for near-extremal PBHs that form within the window relevant for dark matter, and $\nu = 0.36$ for collapse within a radiation-dominated fluid, the leading corrections to $P (\bar{\cal C})$ are suppressed by ${\cal O} (10^{-25})$ compared to the simple Press-Schechter distribution in the limit $m \ll 1$.

To study the abundance of near-extremal PBHs within the regime of interest, we therefore approximate 
\beq
P (\bar{\cal C}) \simeq \frac{ 2 }{ ( \pi \sigma^2 )^{1/2} } \exp \left[ - \frac{ {\cal C}_c^2 }{ (2 \sigma^2)} \right] + {\cal O} (m^{1/\nu}),
\label{PbarC1}
\eeq
where the variance is defined via 
\beq
\sigma^2 = \int d \, {\rm ln } \, k \, W^2 (k, R) \, {\cal P}_{\cal C} (k)
\label{sigmadef}
\eeq
in terms of a window function $W (k, R)$, a smoothing scale $R$, and the dimensionless power spectrum for primordial overdensities ${\cal P}_{\cal C} (k)$ \cite{DeLuca:2019qsy,Gow:2020bzo,Gow:2022jfb}. The compaction has a nonlinear relationship with the gauge-invariant scalar curvature perturbation ${\cal R}$ and its first (radial) derivative. On scales of interest, $k \simeq k_{\rm PBH}$, a conservative estimate is to impose the threshold ${\cal P}_{\cal R} (k_{\rm PBH}) \geq 10^{-3}$ in order for PBHs to form following the end of inflation, when perturbations of comoving wavenumber $k_{\rm PBH}$ cross back inside the Hubble radius. On the other hand, significantly larger perturbations, with ${\cal P}_{\cal R} (k_{\rm PBH}) > 10^{-2}$, would yield an overproduction of PBHs, far in excess of the present dark matter abundance \cite{Kalaja:2019uju,Gow:2020bzo,Gow:2022jfb}. Given the relationship between ${\cal C}$ and ${\cal R}$, these constraints yield a range of interest for the variance of the compaction, $10^{-2} \leq \sigma^2 \leq 10^{-1}$ \cite{DeLuca:2019qsy,Gow:2020bzo,Gow:2022jfb}.


\begin{thebibliography}{100}%
\makeatletter
\providecommand \@ifxundefined [1]{%
 \@ifx{#1\undefined}
}%
\providecommand \@ifnum [1]{%
 \ifnum #1\expandafter \@firstoftwo
 \else \expandafter \@secondoftwo
 \fi
}%
\providecommand \@ifx [1]{%
 \ifx #1\expandafter \@firstoftwo
 \else \expandafter \@secondoftwo
 \fi
}%
\providecommand \natexlab [1]{#1}%
\providecommand \enquote  [1]{``#1''}%
\providecommand \bibnamefont  [1]{#1}%
\providecommand \bibfnamefont [1]{#1}%
\providecommand \citenamefont [1]{#1}%
\providecommand \href@noop [0]{\@secondoftwo}%
\providecommand \href [0]{\begingroup \@sanitize@url \@href}%
\providecommand \@href[1]{\@@startlink{#1}\@@href}%
\providecommand \@@href[1]{\endgroup#1\@@endlink}%
\providecommand \@sanitize@url [0]{\catcode `\\12\catcode `\$12\catcode
  `\&12\catcode `\#12\catcode `\^12\catcode `\_12\catcode `\%12\relax}%
\providecommand \@@startlink[1]{}%
\providecommand \@@endlink[0]{}%
\providecommand \url  [0]{\begingroup\@sanitize@url \@url }%
\providecommand \@url [1]{\endgroup\@href {#1}{\urlprefix }}%
\providecommand \urlprefix  [0]{URL }%
\providecommand \Eprint [0]{\href }%
\providecommand \doibase [0]{http://dx.doi.org/}%
\providecommand \selectlanguage [0]{\@gobble}%
\providecommand \bibinfo  [0]{\@secondoftwo}%
\providecommand \bibfield  [0]{\@secondoftwo}%
\providecommand \translation [1]{[#1]}%
\providecommand \BibitemOpen [0]{}%
\providecommand \bibitemStop [0]{}%
\providecommand \bibitemNoStop [0]{.\EOS\space}%
\providecommand \EOS [0]{\spacefactor3000\relax}%
\providecommand \BibitemShut  [1]{\csname bibitem#1\endcsname}%
\let\auto@bib@innerbib\@empty
\bibitem [{\citenamefont {Zel'dovich}(1967)}]{Zeldovich:1967lct}%
  \BibitemOpen
  \bibfield  {author} {\bibinfo {author} {\bibfnamefont {I.~D.}\ \bibnamefont
  {Zel'dovich}, \bibfnamefont {Ya.B.;~Novikov}},\ }\bibfield  {title} {\enquote
  {\bibinfo {title} {{The Hypothesis of Cores Retarded during Expansion and the
  Hot Cosmological Model}},}\ }\href@noop {} {\bibfield  {journal} {\bibinfo
  {journal} {Soviet Astron. AJ (Engl. Transl. ),}\ }\textbf {\bibinfo {volume}
  {10}},\ \bibinfo {pages} {602} (\bibinfo {year} {1967})}\BibitemShut
  {NoStop}%
\bibitem [{\citenamefont {Hawking}(1971)}]{Hawking:1971ei}%
  \BibitemOpen
  \bibfield  {author} {\bibinfo {author} {\bibfnamefont {Stephen}\ \bibnamefont
  {Hawking}},\ }\bibfield  {title} {\enquote {\bibinfo {title}
  {{Gravitationally collapsed objects of very low mass}},}\ }\href@noop {}
  {\bibfield  {journal} {\bibinfo  {journal} {Mon. Not. Roy. Astron. Soc.}\
  }\textbf {\bibinfo {volume} {152}},\ \bibinfo {pages} {75} (\bibinfo {year}
  {1971})}\BibitemShut {NoStop}%
\bibitem [{\citenamefont {Carr}\ and\ \citenamefont
  {Hawking}(1974)}]{Carr:1974nx}%
  \BibitemOpen
  \bibfield  {author} {\bibinfo {author} {\bibfnamefont {Bernard~J.}\
  \bibnamefont {Carr}}\ and\ \bibinfo {author} {\bibfnamefont {S.~W.}\
  \bibnamefont {Hawking}},\ }\bibfield  {title} {\enquote {\bibinfo {title}
  {{Black holes in the early Universe}},}\ }\href {\doibase
  10.1093/mnras/168.2.399} {\bibfield  {journal} {\bibinfo  {journal} {Mon.
  Not. Roy. Astron. Soc.}\ }\textbf {\bibinfo {volume} {168}},\ \bibinfo
  {pages} {399--415} (\bibinfo {year} {1974})}\BibitemShut {NoStop}%
\bibitem [{\citenamefont {Khlopov}(2010)}]{Khlopov:2008qy}%
  \BibitemOpen
  \bibfield  {author} {\bibinfo {author} {\bibfnamefont {Maxim~Yu.}\
  \bibnamefont {Khlopov}},\ }\bibfield  {title} {\enquote {\bibinfo {title}
  {{Primordial Black Holes}},}\ }\href {\doibase 10.1088/1674-4527/10/6/001}
  {\bibfield  {journal} {\bibinfo  {journal} {Res. Astron. Astrophys.}\
  }\textbf {\bibinfo {volume} {10}},\ \bibinfo {pages} {495--528} (\bibinfo
  {year} {2010})},\ \Eprint {http://arxiv.org/abs/0801.0116} {arXiv:0801.0116
  [astro-ph]} \BibitemShut {NoStop}%
\bibitem [{\citenamefont {{Carr}}\ and\ \citenamefont
  {{K\"{u}hnel}}(2020)}]{Carr:2020xqk}%
  \BibitemOpen
  \bibfield  {author} {\bibinfo {author} {\bibfnamefont {Bernard}\ \bibnamefont
  {{Carr}}}\ and\ \bibinfo {author} {\bibfnamefont {Florian}\ \bibnamefont
  {{K\"{u}hnel}}},\ }\bibfield  {title} {\enquote {\bibinfo {title}
  {{Primordial Black Holes as Dark Matter: Recent Developments}},}\ }\href
  {\doibase 10.1146/annurev-nucl-050520-125911} {\bibfield  {journal} {\bibinfo
   {journal} {Ann. Rev. Nucl. Part. Sci.}\ }\textbf {\bibinfo {volume} {70}},\
  \bibinfo {pages} {355--394} (\bibinfo {year} {2020})},\ \Eprint
  {http://arxiv.org/abs/2006.02838} {arXiv:2006.02838 [astro-ph.CO]}
  \BibitemShut {NoStop}%
\bibitem [{\citenamefont {Green}\ and\ \citenamefont
  {Kavanagh}(2021)}]{Green:2020jor}%
  \BibitemOpen
  \bibfield  {author} {\bibinfo {author} {\bibfnamefont {Anne~M.}\ \bibnamefont
  {Green}}\ and\ \bibinfo {author} {\bibfnamefont {Bradley~J.}\ \bibnamefont
  {Kavanagh}},\ }\bibfield  {title} {\enquote {\bibinfo {title} {{Primordial
  Black Holes as a dark matter candidate}},}\ }\href {\doibase
  10.1088/1361-6471/abc534} {\bibfield  {journal} {\bibinfo  {journal} {J.
  Phys. G}\ }\textbf {\bibinfo {volume} {48}},\ \bibinfo {pages} {043001}
  (\bibinfo {year} {2021})},\ \Eprint {http://arxiv.org/abs/2007.10722}
  {arXiv:2007.10722 [astro-ph.CO]} \BibitemShut {NoStop}%
\bibitem [{\citenamefont {Carr}\ \emph {et~al.}(2021)\citenamefont {Carr},
  \citenamefont {Kohri}, \citenamefont {Sendouda},\ and\ \citenamefont
  {Yokoyama}}]{Carr:2020gox}%
  \BibitemOpen
  \bibfield  {author} {\bibinfo {author} {\bibfnamefont {Bernard}\ \bibnamefont
  {Carr}}, \bibinfo {author} {\bibfnamefont {Kazunori}\ \bibnamefont {Kohri}},
  \bibinfo {author} {\bibfnamefont {Yuuiti}\ \bibnamefont {Sendouda}}, \ and\
  \bibinfo {author} {\bibfnamefont {Jun'ichi}\ \bibnamefont {Yokoyama}},\
  }\bibfield  {title} {\enquote {\bibinfo {title} {{Constraints on primordial
  black holes}},}\ }\href {\doibase 10.1088/1361-6633/ac1e31} {\bibfield
  {journal} {\bibinfo  {journal} {Rept. Prog. Phys.}\ }\textbf {\bibinfo
  {volume} {84}},\ \bibinfo {pages} {116902} (\bibinfo {year} {2021})},\
  \Eprint {http://arxiv.org/abs/2002.12778} {arXiv:2002.12778 [astro-ph.CO]}
  \BibitemShut {NoStop}%
\bibitem [{\citenamefont {Villanueva-Domingo}\ \emph
  {et~al.}(2021)\citenamefont {Villanueva-Domingo}, \citenamefont {Mena},\ and\
  \citenamefont {Palomares-Ruiz}}]{Villanueva-Domingo:2021spv}%
  \BibitemOpen
  \bibfield  {author} {\bibinfo {author} {\bibfnamefont {Pablo}\ \bibnamefont
  {Villanueva-Domingo}}, \bibinfo {author} {\bibfnamefont {Olga}\ \bibnamefont
  {Mena}}, \ and\ \bibinfo {author} {\bibfnamefont {Sergio}\ \bibnamefont
  {Palomares-Ruiz}},\ }\bibfield  {title} {\enquote {\bibinfo {title} {{A brief
  review on primordial black holes as dark matter}},}\ }\href {\doibase
  10.3389/fspas.2021.681084} {\bibfield  {journal} {\bibinfo  {journal} {Front.
  Astron. Space Sci.}\ }\textbf {\bibinfo {volume} {8}},\ \bibinfo {pages} {87}
  (\bibinfo {year} {2021})},\ \Eprint {http://arxiv.org/abs/2103.12087}
  {arXiv:2103.12087 [astro-ph.CO]} \BibitemShut {NoStop}%
\bibitem [{\citenamefont {Escriv\`a}\ \emph {et~al.}(2022)\citenamefont
  {Escriv\`a}, \citenamefont {Kuhnel},\ and\ \citenamefont
  {Tada}}]{Escriva:2022duf}%
  \BibitemOpen
  \bibfield  {author} {\bibinfo {author} {\bibfnamefont {Albert}\ \bibnamefont
  {Escriv\`a}}, \bibinfo {author} {\bibfnamefont {Florian}\ \bibnamefont
  {Kuhnel}}, \ and\ \bibinfo {author} {\bibfnamefont {Yuichiro}\ \bibnamefont
  {Tada}},\ }\bibfield  {title} {\enquote {\bibinfo {title} {{Primordial Black
  Holes}},}\ }\href@noop {} {\  (\bibinfo {year} {2022})},\ \Eprint
  {http://arxiv.org/abs/2211.05767} {arXiv:2211.05767 [astro-ph.CO]}
  \BibitemShut {NoStop}%
\bibitem [{\citenamefont {Carr}\ \emph {et~al.}(2024)\citenamefont {Carr},
  \citenamefont {Clesse}, \citenamefont {Garcia-Bellido}, \citenamefont
  {Hawkins},\ and\ \citenamefont {{K\"{u}hnel}}}]{Carr:2023tpt}%
  \BibitemOpen
  \bibfield  {author} {\bibinfo {author} {\bibfnamefont {Bernard}\ \bibnamefont
  {Carr}}, \bibinfo {author} {\bibfnamefont {Sebastien}\ \bibnamefont
  {Clesse}}, \bibinfo {author} {\bibfnamefont {Juan}\ \bibnamefont
  {Garcia-Bellido}}, \bibinfo {author} {\bibfnamefont {Michael}\ \bibnamefont
  {Hawkins}}, \ and\ \bibinfo {author} {\bibfnamefont {Florian}\ \bibnamefont
  {{K\"{u}hnel}}},\ }\bibfield  {title} {\enquote {\bibinfo {title}
  {{Observational evidence for primordial black holes: A positivist
  perspective}},}\ }\href {\doibase 10.1016/j.physrep.2023.11.005} {\bibfield
  {journal} {\bibinfo  {journal} {Phys. Rept.}\ }\textbf {\bibinfo {volume}
  {1054}},\ \bibinfo {pages} {1--68} (\bibinfo {year} {2024})},\ \Eprint
  {http://arxiv.org/abs/2306.03903} {arXiv:2306.03903 [astro-ph.CO]}
  \BibitemShut {NoStop}%
\bibitem [{\citenamefont {Belotsky}\ \emph {et~al.}(2014)\citenamefont
  {Belotsky}, \citenamefont {Dmitriev}, \citenamefont {Esipova}, \citenamefont
  {Gani}, \citenamefont {Grobov}, \citenamefont {Khlopov}, \citenamefont
  {Kirillov}, \citenamefont {Rubin},\ and\ \citenamefont
  {Svadkovsky}}]{Belotsky:2014kca}%
  \BibitemOpen
  \bibfield  {author} {\bibinfo {author} {\bibfnamefont {K.~M.}\ \bibnamefont
  {Belotsky}}, \bibinfo {author} {\bibfnamefont {A.~D.}\ \bibnamefont
  {Dmitriev}}, \bibinfo {author} {\bibfnamefont {E.~A.}\ \bibnamefont
  {Esipova}}, \bibinfo {author} {\bibfnamefont {V.~A.}\ \bibnamefont {Gani}},
  \bibinfo {author} {\bibfnamefont {A.~V.}\ \bibnamefont {Grobov}}, \bibinfo
  {author} {\bibfnamefont {M.~Yu.}\ \bibnamefont {Khlopov}}, \bibinfo {author}
  {\bibfnamefont {A.~A.}\ \bibnamefont {Kirillov}}, \bibinfo {author}
  {\bibfnamefont {S.~G.}\ \bibnamefont {Rubin}}, \ and\ \bibinfo {author}
  {\bibfnamefont {I.~V.}\ \bibnamefont {Svadkovsky}},\ }\bibfield  {title}
  {\enquote {\bibinfo {title} {{Signatures of primordial black hole dark
  matter}},}\ }\href {\doibase 10.1142/S0217732314400057} {\bibfield  {journal}
  {\bibinfo  {journal} {Mod. Phys. Lett. A}\ }\textbf {\bibinfo {volume}
  {29}},\ \bibinfo {pages} {1440005} (\bibinfo {year} {2014})},\ \Eprint
  {http://arxiv.org/abs/1410.0203} {arXiv:1410.0203 [astro-ph.CO]} \BibitemShut
  {NoStop}%
\bibitem [{\citenamefont {\"Ozsoy}\ and\ \citenamefont
  {Tasinato}(2023)}]{Ozsoy:2023ryl}%
  \BibitemOpen
  \bibfield  {author} {\bibinfo {author} {\bibfnamefont {Ogan}\ \bibnamefont
  {\"Ozsoy}}\ and\ \bibinfo {author} {\bibfnamefont {Gianmassimo}\ \bibnamefont
  {Tasinato}},\ }\bibfield  {title} {\enquote {\bibinfo {title} {{Inflation and
  Primordial Black Holes}},}\ }\href@noop {} {\  (\bibinfo {year} {2023})},\
  \Eprint {http://arxiv.org/abs/2301.03600} {arXiv:2301.03600 [astro-ph.CO]}
  \BibitemShut {NoStop}%
\bibitem [{\citenamefont {Carr}(1975)}]{Carr:1975qj}%
  \BibitemOpen
  \bibfield  {author} {\bibinfo {author} {\bibfnamefont {Bernard~J.}\
  \bibnamefont {Carr}},\ }\bibfield  {title} {\enquote {\bibinfo {title} {{The
  Primordial black hole mass spectrum}},}\ }\href {\doibase 10.1086/153853}
  {\bibfield  {journal} {\bibinfo  {journal} {Astrophys. J.}\ }\textbf
  {\bibinfo {volume} {201}},\ \bibinfo {pages} {1--19} (\bibinfo {year}
  {1975})}\BibitemShut {NoStop}%
\bibitem [{\citenamefont {Niemeyer}\ and\ \citenamefont
  {Jedamzik}(1998)}]{Niemeyer:1997mt}%
  \BibitemOpen
  \bibfield  {author} {\bibinfo {author} {\bibfnamefont {Jens~C.}\ \bibnamefont
  {Niemeyer}}\ and\ \bibinfo {author} {\bibfnamefont {K.}~\bibnamefont
  {Jedamzik}},\ }\bibfield  {title} {\enquote {\bibinfo {title} {{Near-critical
  gravitational collapse and the initial mass function of primordial black
  holes}},}\ }\href {\doibase 10.1103/PhysRevLett.80.5481} {\bibfield
  {journal} {\bibinfo  {journal} {Phys. Rev. Lett.}\ }\textbf {\bibinfo
  {volume} {80}},\ \bibinfo {pages} {5481--5484} (\bibinfo {year} {1998})},\
  \Eprint {http://arxiv.org/abs/astro-ph/9709072} {arXiv:astro-ph/9709072}
  \BibitemShut {NoStop}%
\bibitem [{\citenamefont {Green}\ and\ \citenamefont
  {Liddle}(1999)}]{Green:1999xm}%
  \BibitemOpen
  \bibfield  {author} {\bibinfo {author} {\bibfnamefont {Anne~M.}\ \bibnamefont
  {Green}}\ and\ \bibinfo {author} {\bibfnamefont {Andrew~R.}\ \bibnamefont
  {Liddle}},\ }\bibfield  {title} {\enquote {\bibinfo {title} {{Critical
  collapse and the primordial black hole initial mass function}},}\ }\href
  {\doibase 10.1103/PhysRevD.60.063509} {\bibfield  {journal} {\bibinfo
  {journal} {Phys. Rev. D}\ }\textbf {\bibinfo {volume} {60}},\ \bibinfo
  {pages} {063509} (\bibinfo {year} {1999})},\ \Eprint
  {http://arxiv.org/abs/astro-ph/9901268} {arXiv:astro-ph/9901268} \BibitemShut
  {NoStop}%
\bibitem [{\citenamefont {K\"uhnel}\ \emph {et~al.}(2016)\citenamefont
  {K\"uhnel}, \citenamefont {Rampf},\ and\ \citenamefont
  {Sandstad}}]{Kuhnel:2015vtw}%
  \BibitemOpen
  \bibfield  {author} {\bibinfo {author} {\bibfnamefont {Florian}\ \bibnamefont
  {K\"uhnel}}, \bibinfo {author} {\bibfnamefont {Cornelius}\ \bibnamefont
  {Rampf}}, \ and\ \bibinfo {author} {\bibfnamefont {Marit}\ \bibnamefont
  {Sandstad}},\ }\bibfield  {title} {\enquote {\bibinfo {title} {{Effects of
  Critical Collapse on Primordial Black-Hole Mass Spectra}},}\ }\href {\doibase
  10.1140/epjc/s10052-016-3945-8} {\bibfield  {journal} {\bibinfo  {journal}
  {Eur. Phys. J. C}\ }\textbf {\bibinfo {volume} {76}},\ \bibinfo {pages} {93}
  (\bibinfo {year} {2016})},\ \Eprint {http://arxiv.org/abs/1512.00488}
  {arXiv:1512.00488 [astro-ph.CO]} \BibitemShut {NoStop}%
\bibitem [{\citenamefont {Mukhanov}(2005)}]{Mukhanov:2005sc}%
  \BibitemOpen
  \bibfield  {author} {\bibinfo {author} {\bibfnamefont {V.}~\bibnamefont
  {Mukhanov}},\ }\href {\doibase 10.1017/CBO9780511790553} {\emph {\bibinfo
  {title} {{Physical Foundations of Cosmology}}}}\ (\bibinfo  {publisher}
  {Cambridge University Press},\ \bibinfo {address} {New York},\ \bibinfo
  {year} {2005})\BibitemShut {NoStop}%
\bibitem [{\citenamefont {Boyanovsky}\ \emph {et~al.}(2006)\citenamefont
  {Boyanovsky}, \citenamefont {de~Vega},\ and\ \citenamefont
  {Schwarz}}]{Boyanovsky:2006bf}%
  \BibitemOpen
  \bibfield  {author} {\bibinfo {author} {\bibfnamefont {D.}~\bibnamefont
  {Boyanovsky}}, \bibinfo {author} {\bibfnamefont {H.~J.}\ \bibnamefont
  {de~Vega}}, \ and\ \bibinfo {author} {\bibfnamefont {D.~J.}\ \bibnamefont
  {Schwarz}},\ }\bibfield  {title} {\enquote {\bibinfo {title} {{Phase
  transitions in the early and the present universe}},}\ }\href {\doibase
  10.1146/annurev.nucl.56.080805.140539} {\bibfield  {journal} {\bibinfo
  {journal} {Ann. Rev. Nucl. Part. Sci.}\ }\textbf {\bibinfo {volume} {56}},\
  \bibinfo {pages} {441--500} (\bibinfo {year} {2006})},\ \Eprint
  {http://arxiv.org/abs/hep-ph/0602002} {arXiv:hep-ph/0602002} \BibitemShut
  {NoStop}%
\bibitem [{\citenamefont {Kapusta}\ and\ \citenamefont
  {Gale}(2006)}]{Kapusta:2006pm}%
  \BibitemOpen
  \bibfield  {author} {\bibinfo {author} {\bibfnamefont {J.~I.}\ \bibnamefont
  {Kapusta}}\ and\ \bibinfo {author} {\bibfnamefont {Charles}\ \bibnamefont
  {Gale}},\ }\href {\doibase 10.1017/CBO9780511535130} {\emph {\bibinfo {title}
  {{Finite-Temperature Field Theory: Principles and Applications}}}}\ (\bibinfo
   {publisher} {Cambridge University Press},\ \bibinfo {address} {Cambridge,
  UK},\ \bibinfo {year} {2006})\BibitemShut {NoStop}%
\bibitem [{\citenamefont {Blaizot}\ and\ \citenamefont
  {Iancu}(2002)}]{Blaizot:2001nr}%
  \BibitemOpen
  \bibfield  {author} {\bibinfo {author} {\bibfnamefont {Jean-Paul}\
  \bibnamefont {Blaizot}}\ and\ \bibinfo {author} {\bibfnamefont {Edmond}\
  \bibnamefont {Iancu}},\ }\bibfield  {title} {\enquote {\bibinfo {title} {{The
  quark gluon plasma: Collective dynamics and hard thermal loops}},}\ }\href
  {\doibase 10.1016/S0370-1573(01)00061-8} {\bibfield  {journal} {\bibinfo
  {journal} {Phys. Rept.}\ }\textbf {\bibinfo {volume} {359}},\ \bibinfo
  {pages} {355--528} (\bibinfo {year} {2002})},\ \Eprint
  {http://arxiv.org/abs/hep-ph/0101103} {arXiv:hep-ph/0101103} \BibitemShut
  {NoStop}%
\bibitem [{\citenamefont {Litim}\ and\ \citenamefont
  {Manuel}(2002)}]{Litim:2001db}%
  \BibitemOpen
  \bibfield  {author} {\bibinfo {author} {\bibfnamefont {Daniel~F.}\
  \bibnamefont {Litim}}\ and\ \bibinfo {author} {\bibfnamefont {Cristina}\
  \bibnamefont {Manuel}},\ }\bibfield  {title} {\enquote {\bibinfo {title}
  {{Semiclassical transport theory for nonAbelian plasmas}},}\ }\href {\doibase
  10.1016/S0370-1573(02)00015-7} {\bibfield  {journal} {\bibinfo  {journal}
  {Phys. Rept.}\ }\textbf {\bibinfo {volume} {364}},\ \bibinfo {pages}
  {451--539} (\bibinfo {year} {2002})},\ \Eprint
  {http://arxiv.org/abs/hep-ph/0110104} {arXiv:hep-ph/0110104} \BibitemShut
  {NoStop}%
\bibitem [{\citenamefont {Mrowczynski}\ \emph {et~al.}(2017)\citenamefont
  {Mrowczynski}, \citenamefont {Schenke},\ and\ \citenamefont
  {Strickland}}]{Mrowczynski:2016etf}%
  \BibitemOpen
  \bibfield  {author} {\bibinfo {author} {\bibfnamefont {Stanislaw}\
  \bibnamefont {Mrowczynski}}, \bibinfo {author} {\bibfnamefont {Bjoern}\
  \bibnamefont {Schenke}}, \ and\ \bibinfo {author} {\bibfnamefont {Michael}\
  \bibnamefont {Strickland}},\ }\bibfield  {title} {\enquote {\bibinfo {title}
  {{Color instabilities in the quark\textendash{}gluon plasma}},}\ }\href
  {\doibase 10.1016/j.physrep.2017.03.003} {\bibfield  {journal} {\bibinfo
  {journal} {Phys. Rept.}\ }\textbf {\bibinfo {volume} {682}},\ \bibinfo
  {pages} {1--97} (\bibinfo {year} {2017})},\ \Eprint
  {http://arxiv.org/abs/1603.08946} {arXiv:1603.08946 [hep-ph]} \BibitemShut
  {NoStop}%
\bibitem [{\citenamefont {Manuel}\ and\ \citenamefont
  {Mrowczynski}(2003)}]{Manuel:2003zr}%
  \BibitemOpen
  \bibfield  {author} {\bibinfo {author} {\bibfnamefont {Cristina}\
  \bibnamefont {Manuel}}\ and\ \bibinfo {author} {\bibfnamefont {Stanislaw}\
  \bibnamefont {Mrowczynski}},\ }\bibfield  {title} {\enquote {\bibinfo {title}
  {{Local equilibrium of the quark gluon plasma}},}\ }\href {\doibase
  10.1103/PhysRevD.68.094010} {\bibfield  {journal} {\bibinfo  {journal} {Phys.
  Rev. D}\ }\textbf {\bibinfo {volume} {68}},\ \bibinfo {pages} {094010}
  (\bibinfo {year} {2003})},\ \Eprint {http://arxiv.org/abs/hep-ph/0306209}
  {arXiv:hep-ph/0306209} \BibitemShut {NoStop}%
\bibitem [{\citenamefont {Manuel}\ and\ \citenamefont
  {Mrowczynski}(2004)}]{Manuel:2004gk}%
  \BibitemOpen
  \bibfield  {author} {\bibinfo {author} {\bibfnamefont {Cristina}\
  \bibnamefont {Manuel}}\ and\ \bibinfo {author} {\bibfnamefont {Stanislaw}\
  \bibnamefont {Mrowczynski}},\ }\bibfield  {title} {\enquote {\bibinfo {title}
  {{Whitening of the quark gluon plasma}},}\ }\href {\doibase
  10.1103/PhysRevD.70.094019} {\bibfield  {journal} {\bibinfo  {journal} {Phys.
  Rev. D}\ }\textbf {\bibinfo {volume} {70}},\ \bibinfo {pages} {094019}
  (\bibinfo {year} {2004})},\ \Eprint {http://arxiv.org/abs/hep-ph/0403024}
  {arXiv:hep-ph/0403024} \BibitemShut {NoStop}%
\bibitem [{\citenamefont {Choptuik}(1993)}]{Choptuik:1992jv}%
  \BibitemOpen
  \bibfield  {author} {\bibinfo {author} {\bibfnamefont {Matthew~W.}\
  \bibnamefont {Choptuik}},\ }\bibfield  {title} {\enquote {\bibinfo {title}
  {{Universality and scaling in gravitational collapse of a massless scalar
  field}},}\ }\href {\doibase 10.1103/PhysRevLett.70.9} {\bibfield  {journal}
  {\bibinfo  {journal} {Phys. Rev. Lett.}\ }\textbf {\bibinfo {volume} {70}},\
  \bibinfo {pages} {9--12} (\bibinfo {year} {1993})}\BibitemShut {NoStop}%
\bibitem [{\citenamefont {Evans}\ and\ \citenamefont
  {Coleman}(1994)}]{Evans:1994pj}%
  \BibitemOpen
  \bibfield  {author} {\bibinfo {author} {\bibfnamefont {Charles~R.}\
  \bibnamefont {Evans}}\ and\ \bibinfo {author} {\bibfnamefont {Jason~S.}\
  \bibnamefont {Coleman}},\ }\bibfield  {title} {\enquote {\bibinfo {title}
  {{Observation of critical phenomena and selfsimilarity in the gravitational
  collapse of radiation fluid}},}\ }\href {\doibase
  10.1103/PhysRevLett.72.1782} {\bibfield  {journal} {\bibinfo  {journal}
  {Phys. Rev. Lett.}\ }\textbf {\bibinfo {volume} {72}},\ \bibinfo {pages}
  {1782--1785} (\bibinfo {year} {1994})},\ \Eprint
  {http://arxiv.org/abs/gr-qc/9402041} {arXiv:gr-qc/9402041} \BibitemShut
  {NoStop}%
\bibitem [{\citenamefont {Niemeyer}\ and\ \citenamefont
  {Jedamzik}(1999)}]{Niemeyer:1999ak}%
  \BibitemOpen
  \bibfield  {author} {\bibinfo {author} {\bibfnamefont {Jens~C.}\ \bibnamefont
  {Niemeyer}}\ and\ \bibinfo {author} {\bibfnamefont {K.}~\bibnamefont
  {Jedamzik}},\ }\bibfield  {title} {\enquote {\bibinfo {title} {{Dynamics of
  primordial black hole formation}},}\ }\href {\doibase
  10.1103/PhysRevD.59.124013} {\bibfield  {journal} {\bibinfo  {journal} {Phys.
  Rev. D}\ }\textbf {\bibinfo {volume} {59}},\ \bibinfo {pages} {124013}
  (\bibinfo {year} {1999})},\ \Eprint {http://arxiv.org/abs/astro-ph/9901292}
  {arXiv:astro-ph/9901292} \BibitemShut {NoStop}%
\bibitem [{\citenamefont {Gundlach}(2003)}]{Gundlach:2002sx}%
  \BibitemOpen
  \bibfield  {author} {\bibinfo {author} {\bibfnamefont {Carsten}\ \bibnamefont
  {Gundlach}},\ }\bibfield  {title} {\enquote {\bibinfo {title} {{Critical
  phenomena in gravitational collapse}},}\ }\href {\doibase
  10.1016/S0370-1573(02)00560-4} {\bibfield  {journal} {\bibinfo  {journal}
  {Phys. Rept.}\ }\textbf {\bibinfo {volume} {376}},\ \bibinfo {pages}
  {339--405} (\bibinfo {year} {2003})},\ \Eprint
  {http://arxiv.org/abs/gr-qc/0210101} {arXiv:gr-qc/0210101} \BibitemShut
  {NoStop}%
\bibitem [{\citenamefont {Gundlach}\ and\ \citenamefont
  {Martin-Garcia}(2007)}]{Gundlach:2007gc}%
  \BibitemOpen
  \bibfield  {author} {\bibinfo {author} {\bibfnamefont {Carsten}\ \bibnamefont
  {Gundlach}}\ and\ \bibinfo {author} {\bibfnamefont {Jose~M.}\ \bibnamefont
  {Martin-Garcia}},\ }\bibfield  {title} {\enquote {\bibinfo {title} {{Critical
  phenomena in gravitational collapse}},}\ }\href {\doibase
  10.12942/lrr-2007-5} {\bibfield  {journal} {\bibinfo  {journal} {Living Rev.
  Rel.}\ }\textbf {\bibinfo {volume} {10}},\ \bibinfo {pages} {5} (\bibinfo
  {year} {2007})},\ \Eprint {http://arxiv.org/abs/0711.4620} {arXiv:0711.4620
  [gr-qc]} \BibitemShut {NoStop}%
\bibitem [{\citenamefont {Musco}\ \emph {et~al.}(2009)\citenamefont {Musco},
  \citenamefont {Miller},\ and\ \citenamefont {Polnarev}}]{Musco:2008hv}%
  \BibitemOpen
  \bibfield  {author} {\bibinfo {author} {\bibfnamefont {Ilia}\ \bibnamefont
  {Musco}}, \bibinfo {author} {\bibfnamefont {John~C.}\ \bibnamefont {Miller}},
  \ and\ \bibinfo {author} {\bibfnamefont {Alexander~G.}\ \bibnamefont
  {Polnarev}},\ }\bibfield  {title} {\enquote {\bibinfo {title} {{Primordial
  black hole formation in the radiative era: Investigation of the critical
  nature of the collapse}},}\ }\href {\doibase 10.1088/0264-9381/26/23/235001}
  {\bibfield  {journal} {\bibinfo  {journal} {Class. Quant. Grav.}\ }\textbf
  {\bibinfo {volume} {26}},\ \bibinfo {pages} {235001} (\bibinfo {year}
  {2009})},\ \Eprint {http://arxiv.org/abs/0811.1452} {arXiv:0811.1452 [gr-qc]}
  \BibitemShut {NoStop}%
\bibitem [{\citenamefont {Escriv\`a}(2022)}]{Escriva:2021aeh}%
  \BibitemOpen
  \bibfield  {author} {\bibinfo {author} {\bibfnamefont {Albert}\ \bibnamefont
  {Escriv\`a}},\ }\bibfield  {title} {\enquote {\bibinfo {title} {{PBH
  Formation from Spherically Symmetric Hydrodynamical Perturbations: A
  Review}},}\ }\href {\doibase 10.3390/universe8020066} {\bibfield  {journal}
  {\bibinfo  {journal} {Universe}\ }\textbf {\bibinfo {volume} {8}},\ \bibinfo
  {pages} {66} (\bibinfo {year} {2022})},\ \Eprint
  {http://arxiv.org/abs/2111.12693} {arXiv:2111.12693 [gr-qc]} \BibitemShut
  {NoStop}%
\bibitem [{\citenamefont {Volkov}\ and\ \citenamefont
  {Gal'tsov}(1999)}]{Volkov:1998cc}%
  \BibitemOpen
  \bibfield  {author} {\bibinfo {author} {\bibfnamefont {Mikhail~S.}\
  \bibnamefont {Volkov}}\ and\ \bibinfo {author} {\bibfnamefont {Dmitri~V.}\
  \bibnamefont {Gal'tsov}},\ }\bibfield  {title} {\enquote {\bibinfo {title}
  {{Gravitating non-Abelian solitons and black holes with Yang-Mills
  fields}},}\ }\href {\doibase 10.1016/S0370-1573(99)00010-1} {\bibfield
  {journal} {\bibinfo  {journal} {Phys. Rept.}\ }\textbf {\bibinfo {volume}
  {319}},\ \bibinfo {pages} {1--83} (\bibinfo {year} {1999})},\ \Eprint
  {http://arxiv.org/abs/hep-th/9810070} {arXiv:hep-th/9810070} \BibitemShut
  {NoStop}%
\bibitem [{\citenamefont {Volkov}(2017)}]{Volkov:2016ehx}%
  \BibitemOpen
  \bibfield  {author} {\bibinfo {author} {\bibfnamefont {Mikhail~S.}\
  \bibnamefont {Volkov}},\ }\bibfield  {title} {\enquote {\bibinfo {title}
  {{Hairy black holes in the XX-th and XXI-st centuries}},}\ }in\ \href
  {\doibase 10.1142/9789813226609_0184} {\emph {\bibinfo {booktitle} {{14th
  Marcel Grossmann Meeting on Recent Developments in Theoretical and
  Experimental General Relativity, Astrophysics, and Relativistic Field
  Theories}}}},\ Vol.~\bibinfo {volume} {2}\ (\bibinfo {year} {2017})\ pp.\
  \bibinfo {pages} {1779--1798},\ \Eprint {http://arxiv.org/abs/1601.08230}
  {arXiv:1601.08230 [gr-qc]} \BibitemShut {NoStop}%
\bibitem [{\citenamefont {Pasechnik}\ and\ \citenamefont
  {\v{S}umbera}(2017)}]{Pasechnik:2016wkt}%
  \BibitemOpen
  \bibfield  {author} {\bibinfo {author} {\bibfnamefont {Roman}\ \bibnamefont
  {Pasechnik}}\ and\ \bibinfo {author} {\bibfnamefont {Michal}\ \bibnamefont
  {\v{S}umbera}},\ }\bibfield  {title} {\enquote {\bibinfo {title}
  {{Phenomenological Review on Quark\textendash{}Gluon Plasma: Concepts vs.
  Observations}},}\ }\href {\doibase 10.3390/universe3010007} {\bibfield
  {journal} {\bibinfo  {journal} {Universe}\ }\textbf {\bibinfo {volume} {3}},\
  \bibinfo {pages} {7} (\bibinfo {year} {2017})},\ \Eprint
  {http://arxiv.org/abs/1611.01533} {arXiv:1611.01533 [hep-ph]} \BibitemShut
  {NoStop}%
\bibitem [{Uni()}]{UnitsNote}%
  \BibitemOpen
  \href@noop {} {}\bibinfo {note} {In this paper, we adopt ``natural'' units
  $\hbar = c = k_B = 1$, restrict attention to $(3 + 1)$ spacetime dimensions,
  and work in terms of the reduced Planck mass, $M_{\rm pl} \equiv 1 / \sqrt{ 8
  \pi G} \simeq 2.43 \times 10^{18} \, {\rm GeV}$.}\BibitemShut {Stop}%
\bibitem [{\citenamefont {Shibata}\ and\ \citenamefont
  {Sasaki}(1999)}]{Shibata:1999zs}%
  \BibitemOpen
  \bibfield  {author} {\bibinfo {author} {\bibfnamefont {Masaru}\ \bibnamefont
  {Shibata}}\ and\ \bibinfo {author} {\bibfnamefont {Misao}\ \bibnamefont
  {Sasaki}},\ }\bibfield  {title} {\enquote {\bibinfo {title} {{Black hole
  formation in the Friedmann universe: Formulation and computation in numerical
  relativity}},}\ }\href {\doibase 10.1103/PhysRevD.60.084002} {\bibfield
  {journal} {\bibinfo  {journal} {Phys. Rev. D}\ }\textbf {\bibinfo {volume}
  {60}},\ \bibinfo {pages} {084002} (\bibinfo {year} {1999})},\ \Eprint
  {http://arxiv.org/abs/gr-qc/9905064} {arXiv:gr-qc/9905064} \BibitemShut
  {NoStop}%
\bibitem [{\citenamefont {Harada}\ \emph {et~al.}(2023)\citenamefont {Harada},
  \citenamefont {Yoo},\ and\ \citenamefont {Koga}}]{Harada:2023ffo}%
  \BibitemOpen
  \bibfield  {author} {\bibinfo {author} {\bibfnamefont {Tomohiro}\
  \bibnamefont {Harada}}, \bibinfo {author} {\bibfnamefont {Chul-Moon}\
  \bibnamefont {Yoo}}, \ and\ \bibinfo {author} {\bibfnamefont {Yasutaka}\
  \bibnamefont {Koga}},\ }\bibfield  {title} {\enquote {\bibinfo {title}
  {{Revisiting compaction functions for primordial black hole formation}},}\
  }\href {\doibase 10.1103/PhysRevD.108.043515} {\bibfield  {journal} {\bibinfo
   {journal} {Phys. Rev. D}\ }\textbf {\bibinfo {volume} {108}},\ \bibinfo
  {pages} {043515} (\bibinfo {year} {2023})},\ \Eprint
  {http://arxiv.org/abs/2304.13284} {arXiv:2304.13284 [gr-qc]} \BibitemShut
  {NoStop}%
\bibitem [{\citenamefont {Escriv\`a}\ \emph {et~al.}(2020)\citenamefont
  {Escriv\`a}, \citenamefont {Germani},\ and\ \citenamefont
  {Sheth}}]{Escriva:2019phb}%
  \BibitemOpen
  \bibfield  {author} {\bibinfo {author} {\bibfnamefont {Albert}\ \bibnamefont
  {Escriv\`a}}, \bibinfo {author} {\bibfnamefont {Cristiano}\ \bibnamefont
  {Germani}}, \ and\ \bibinfo {author} {\bibfnamefont {Ravi~K.}\ \bibnamefont
  {Sheth}},\ }\bibfield  {title} {\enquote {\bibinfo {title} {{Universal
  threshold for primordial black hole formation}},}\ }\href {\doibase
  10.1103/PhysRevD.101.044022} {\bibfield  {journal} {\bibinfo  {journal}
  {Phys. Rev. D}\ }\textbf {\bibinfo {volume} {101}},\ \bibinfo {pages}
  {044022} (\bibinfo {year} {2020})},\ \Eprint
  {http://arxiv.org/abs/1907.13311} {arXiv:1907.13311 [gr-qc]} \BibitemShut
  {NoStop}%
\bibitem [{\citenamefont {Musco}(2019)}]{Musco:2018rwt}%
  \BibitemOpen
  \bibfield  {author} {\bibinfo {author} {\bibfnamefont {Ilia}\ \bibnamefont
  {Musco}},\ }\bibfield  {title} {\enquote {\bibinfo {title} {{Threshold for
  primordial black holes: Dependence on the shape of the cosmological
  perturbations}},}\ }\href {\doibase 10.1103/PhysRevD.100.123524} {\bibfield
  {journal} {\bibinfo  {journal} {Phys. Rev. D}\ }\textbf {\bibinfo {volume}
  {100}},\ \bibinfo {pages} {123524} (\bibinfo {year} {2019})},\ \Eprint
  {http://arxiv.org/abs/1809.02127} {arXiv:1809.02127 [gr-qc]} \BibitemShut
  {NoStop}%
\bibitem [{\citenamefont {Ando}\ \emph {et~al.}(2018)\citenamefont {Ando},
  \citenamefont {Inomata},\ and\ \citenamefont {Kawasaki}}]{Ando:2018qdb}%
  \BibitemOpen
  \bibfield  {author} {\bibinfo {author} {\bibfnamefont {Kenta}\ \bibnamefont
  {Ando}}, \bibinfo {author} {\bibfnamefont {Keisuke}\ \bibnamefont {Inomata}},
  \ and\ \bibinfo {author} {\bibfnamefont {Masahiro}\ \bibnamefont
  {Kawasaki}},\ }\bibfield  {title} {\enquote {\bibinfo {title} {{Primordial
  black holes and uncertainties in the choice of the window function}},}\
  }\href {\doibase 10.1103/PhysRevD.97.103528} {\bibfield  {journal} {\bibinfo
  {journal} {Phys. Rev. D}\ }\textbf {\bibinfo {volume} {97}},\ \bibinfo
  {pages} {103528} (\bibinfo {year} {2018})},\ \Eprint
  {http://arxiv.org/abs/1802.06393} {arXiv:1802.06393 [astro-ph.CO]}
  \BibitemShut {NoStop}%
\bibitem [{\citenamefont {Germani}\ and\ \citenamefont
  {Musco}(2019)}]{Germani:2018jgr}%
  \BibitemOpen
  \bibfield  {author} {\bibinfo {author} {\bibfnamefont {Cristiano}\
  \bibnamefont {Germani}}\ and\ \bibinfo {author} {\bibfnamefont {Ilia}\
  \bibnamefont {Musco}},\ }\bibfield  {title} {\enquote {\bibinfo {title}
  {{Abundance of Primordial Black Holes Depends on the Shape of the
  Inflationary Power Spectrum}},}\ }\href {\doibase
  10.1103/PhysRevLett.122.141302} {\bibfield  {journal} {\bibinfo  {journal}
  {Phys. Rev. Lett.}\ }\textbf {\bibinfo {volume} {122}},\ \bibinfo {pages}
  {141302} (\bibinfo {year} {2019})},\ \Eprint
  {http://arxiv.org/abs/1805.04087} {arXiv:1805.04087 [astro-ph.CO]}
  \BibitemShut {NoStop}%
\bibitem [{\citenamefont {Kalaja}\ \emph {et~al.}(2019)\citenamefont {Kalaja},
  \citenamefont {Bellomo}, \citenamefont {Bartolo}, \citenamefont {Bertacca},
  \citenamefont {Matarrese}, \citenamefont {Musco}, \citenamefont
  {Raccanelli},\ and\ \citenamefont {Verde}}]{Kalaja:2019uju}%
  \BibitemOpen
  \bibfield  {author} {\bibinfo {author} {\bibfnamefont {Alba}\ \bibnamefont
  {Kalaja}}, \bibinfo {author} {\bibfnamefont {Nicola}\ \bibnamefont
  {Bellomo}}, \bibinfo {author} {\bibfnamefont {Nicola}\ \bibnamefont
  {Bartolo}}, \bibinfo {author} {\bibfnamefont {Daniele}\ \bibnamefont
  {Bertacca}}, \bibinfo {author} {\bibfnamefont {Sabino}\ \bibnamefont
  {Matarrese}}, \bibinfo {author} {\bibfnamefont {Ilia}\ \bibnamefont {Musco}},
  \bibinfo {author} {\bibfnamefont {Alvise}\ \bibnamefont {Raccanelli}}, \ and\
  \bibinfo {author} {\bibfnamefont {Licia}\ \bibnamefont {Verde}},\ }\bibfield
  {title} {\enquote {\bibinfo {title} {{From Primordial Black Holes Abundance
  to Primordial Curvature Power Spectrum (and back)}},}\ }\href {\doibase
  10.1088/1475-7516/2019/10/031} {\bibfield  {journal} {\bibinfo  {journal}
  {JCAP}\ }\textbf {\bibinfo {volume} {10}},\ \bibinfo {pages} {031} (\bibinfo
  {year} {2019})},\ \Eprint {http://arxiv.org/abs/1908.03596} {arXiv:1908.03596
  [astro-ph.CO]} \BibitemShut {NoStop}%
\bibitem [{\citenamefont {Escriv\`a}(2020)}]{Escriva:2019nsa}%
  \BibitemOpen
  \bibfield  {author} {\bibinfo {author} {\bibfnamefont {Albert}\ \bibnamefont
  {Escriv\`a}},\ }\bibfield  {title} {\enquote {\bibinfo {title} {{Simulation
  of primordial black hole formation using pseudo-spectral methods}},}\ }\href
  {\doibase 10.1016/j.dark.2020.100466} {\bibfield  {journal} {\bibinfo
  {journal} {Phys. Dark Univ.}\ }\textbf {\bibinfo {volume} {27}},\ \bibinfo
  {pages} {100466} (\bibinfo {year} {2020})},\ \Eprint
  {http://arxiv.org/abs/1907.13065} {arXiv:1907.13065 [gr-qc]} \BibitemShut
  {NoStop}%
\bibitem [{\citenamefont {Young}(2019)}]{Young:2019osy}%
  \BibitemOpen
  \bibfield  {author} {\bibinfo {author} {\bibfnamefont {Sam}\ \bibnamefont
  {Young}},\ }\bibfield  {title} {\enquote {\bibinfo {title} {{The primordial
  black hole formation criterion re-examined: Parametrisation, timing and the
  choice of window function}},}\ }\href {\doibase 10.1142/S0218271820300025}
  {\bibfield  {journal} {\bibinfo  {journal} {Int. J. Mod. Phys. D}\ }\textbf
  {\bibinfo {volume} {29}},\ \bibinfo {pages} {2030002} (\bibinfo {year}
  {2019})},\ \Eprint {http://arxiv.org/abs/1905.01230} {arXiv:1905.01230
  [astro-ph.CO]} \BibitemShut {NoStop}%
\bibitem [{\citenamefont {Gow}\ \emph {et~al.}(2021)\citenamefont {Gow},
  \citenamefont {Byrnes}, \citenamefont {Cole},\ and\ \citenamefont
  {Young}}]{Gow:2020bzo}%
  \BibitemOpen
  \bibfield  {author} {\bibinfo {author} {\bibfnamefont {Andrew~D.}\
  \bibnamefont {Gow}}, \bibinfo {author} {\bibfnamefont {Christian~T.}\
  \bibnamefont {Byrnes}}, \bibinfo {author} {\bibfnamefont {Philippa~S.}\
  \bibnamefont {Cole}}, \ and\ \bibinfo {author} {\bibfnamefont {Sam}\
  \bibnamefont {Young}},\ }\bibfield  {title} {\enquote {\bibinfo {title} {{The
  power spectrum on small scales: Robust constraints and comparing PBH
  methodologies}},}\ }\href {\doibase 10.1088/1475-7516/2021/02/002} {\bibfield
   {journal} {\bibinfo  {journal} {JCAP}\ }\textbf {\bibinfo {volume} {02}},\
  \bibinfo {pages} {002} (\bibinfo {year} {2021})},\ \Eprint
  {http://arxiv.org/abs/2008.03289} {arXiv:2008.03289 [astro-ph.CO]}
  \BibitemShut {NoStop}%
\bibitem [{\citenamefont {Koike}\ \emph {et~al.}(1995)\citenamefont {Koike},
  \citenamefont {Hara},\ and\ \citenamefont {Adachi}}]{Koike:1995jm}%
  \BibitemOpen
  \bibfield  {author} {\bibinfo {author} {\bibfnamefont {Tatsuhiko}\
  \bibnamefont {Koike}}, \bibinfo {author} {\bibfnamefont {Takashi}\
  \bibnamefont {Hara}}, \ and\ \bibinfo {author} {\bibfnamefont {Satoshi}\
  \bibnamefont {Adachi}},\ }\bibfield  {title} {\enquote {\bibinfo {title}
  {{Critical behavior in gravitational collapse of radiation fluid: A
  Renormalization group (linear perturbation) analysis}},}\ }\href {\doibase
  10.1103/PhysRevLett.74.5170} {\bibfield  {journal} {\bibinfo  {journal}
  {Phys. Rev. Lett.}\ }\textbf {\bibinfo {volume} {74}},\ \bibinfo {pages}
  {5170--5173} (\bibinfo {year} {1995})},\ \Eprint
  {http://arxiv.org/abs/gr-qc/9503007} {arXiv:gr-qc/9503007} \BibitemShut
  {NoStop}%
\bibitem [{\citenamefont {Choptuik}\ \emph {et~al.}(1996)\citenamefont
  {Choptuik}, \citenamefont {Chmaj},\ and\ \citenamefont
  {Bizon}}]{Choptuik:1996yg}%
  \BibitemOpen
  \bibfield  {author} {\bibinfo {author} {\bibfnamefont {Matthew~W.}\
  \bibnamefont {Choptuik}}, \bibinfo {author} {\bibfnamefont {Tadeusz}\
  \bibnamefont {Chmaj}}, \ and\ \bibinfo {author} {\bibfnamefont {Piotr}\
  \bibnamefont {Bizon}},\ }\bibfield  {title} {\enquote {\bibinfo {title}
  {{Critical behavior in gravitational collapse of a Yang-Mills field}},}\
  }\href {\doibase 10.1103/PhysRevLett.77.424} {\bibfield  {journal} {\bibinfo
  {journal} {Phys. Rev. Lett.}\ }\textbf {\bibinfo {volume} {77}},\ \bibinfo
  {pages} {424--427} (\bibinfo {year} {1996})},\ \Eprint
  {http://arxiv.org/abs/gr-qc/9603051} {arXiv:gr-qc/9603051} \BibitemShut
  {NoStop}%
\bibitem [{\citenamefont {Neilsen}\ and\ \citenamefont
  {Choptuik}(2000)}]{Neilsen:1998qc}%
  \BibitemOpen
  \bibfield  {author} {\bibinfo {author} {\bibfnamefont {David~W.}\
  \bibnamefont {Neilsen}}\ and\ \bibinfo {author} {\bibfnamefont {Matthew~W.}\
  \bibnamefont {Choptuik}},\ }\bibfield  {title} {\enquote {\bibinfo {title}
  {{Critical phenomena in perfect fluids}},}\ }\href {\doibase
  10.1088/0264-9381/17/4/303} {\bibfield  {journal} {\bibinfo  {journal}
  {Class. Quant. Grav.}\ }\textbf {\bibinfo {volume} {17}},\ \bibinfo {pages}
  {761--782} (\bibinfo {year} {2000})},\ \Eprint
  {http://arxiv.org/abs/gr-qc/9812053} {arXiv:gr-qc/9812053} \BibitemShut
  {NoStop}%
\bibitem [{\citenamefont {Choptuik}\ \emph {et~al.}(1999)\citenamefont
  {Choptuik}, \citenamefont {Hirschmann},\ and\ \citenamefont
  {Marsa}}]{Choptuik:1999gh}%
  \BibitemOpen
  \bibfield  {author} {\bibinfo {author} {\bibfnamefont {Matthew~W.}\
  \bibnamefont {Choptuik}}, \bibinfo {author} {\bibfnamefont {Eric~W.}\
  \bibnamefont {Hirschmann}}, \ and\ \bibinfo {author} {\bibfnamefont
  {Robert~L.}\ \bibnamefont {Marsa}},\ }\bibfield  {title} {\enquote {\bibinfo
  {title} {{New critical behavior in Einstein-Yang-Mills collapse}},}\ }\href
  {\doibase 10.1103/PhysRevD.60.124011} {\bibfield  {journal} {\bibinfo
  {journal} {Phys. Rev. D}\ }\textbf {\bibinfo {volume} {60}},\ \bibinfo
  {pages} {124011} (\bibinfo {year} {1999})},\ \Eprint
  {http://arxiv.org/abs/gr-qc/9903081} {arXiv:gr-qc/9903081} \BibitemShut
  {NoStop}%
\bibitem [{Typ()}]{TypeIIb}%
  \BibitemOpen
  \href@noop {} {}\bibinfo {note} {Recent work has identified rare scenarios in
  which primordial curvature perturbations with highly non-Gaussian statistics,
  and with amplitudes far greater than the threshold for collapse, can yield
  departures from the typical expectation for the scaling of the black hole
  mass $M$ with the order parameter $\vert \bar{\cal C} - {\cal C}_c \vert$
  \cite{Kopp:2010sh,Carr:2014pga,Uehara:2024yyp}. As we will see below, in this
  work we are concerned with PBHs that form in a different regime: the
  small-mass tail of the distribution, with $\bar{\cal C} \gtrsim {\cal C}_c$
  rather than $\bar{\cal C} \gg {\cal C}_c$, for which the non-Gaussian
  features remain subdominant.}\BibitemShut {Stop}%
\bibitem [{\citenamefont {Rice}\ and\ \citenamefont
  {Zhang}(2017)}]{Rice:2017avg}%
  \BibitemOpen
  \bibfield  {author} {\bibinfo {author} {\bibfnamefont {Jared~R.}\
  \bibnamefont {Rice}}\ and\ \bibinfo {author} {\bibfnamefont {Bing}\
  \bibnamefont {Zhang}},\ }\bibfield  {title} {\enquote {\bibinfo {title}
  {{Cosmological evolution of primordial black holes}},}\ }\href {\doibase
  10.1016/j.jheap.2017.02.002} {\bibfield  {journal} {\bibinfo  {journal}
  {JHEAp}\ }\textbf {\bibinfo {volume} {13-14}},\ \bibinfo {pages} {22--31}
  (\bibinfo {year} {2017})},\ \Eprint {http://arxiv.org/abs/1702.08069}
  {arXiv:1702.08069 [astro-ph.HE]} \BibitemShut {NoStop}%
\bibitem [{\citenamefont {De~Luca}\ \emph
  {et~al.}(2020{\natexlab{a}})\citenamefont {De~Luca}, \citenamefont
  {Franciolini}, \citenamefont {Pani},\ and\ \citenamefont
  {Riotto}}]{DeLuca:2020fpg}%
  \BibitemOpen
  \bibfield  {author} {\bibinfo {author} {\bibfnamefont {V.}~\bibnamefont
  {De~Luca}}, \bibinfo {author} {\bibfnamefont {G.}~\bibnamefont
  {Franciolini}}, \bibinfo {author} {\bibfnamefont {P.}~\bibnamefont {Pani}}, \
  and\ \bibinfo {author} {\bibfnamefont {A.}~\bibnamefont {Riotto}},\
  }\bibfield  {title} {\enquote {\bibinfo {title} {{Constraints on Primordial
  Black Holes: the Importance of Accretion}},}\ }\href {\doibase
  10.1103/PhysRevD.102.043505} {\bibfield  {journal} {\bibinfo  {journal}
  {Phys. Rev. D}\ }\textbf {\bibinfo {volume} {102}},\ \bibinfo {pages}
  {043505} (\bibinfo {year} {2020}{\natexlab{a}})},\ \Eprint
  {http://arxiv.org/abs/2003.12589} {arXiv:2003.12589 [astro-ph.CO]}
  \BibitemShut {NoStop}%
\bibitem [{\citenamefont {Alonso-Monsalve}\ and\ \citenamefont
  {Kaiser}(2023)}]{Alonso-Monsalve:2023jfq}%
  \BibitemOpen
  \bibfield  {author} {\bibinfo {author} {\bibfnamefont {Elba}\ \bibnamefont
  {Alonso-Monsalve}}\ and\ \bibinfo {author} {\bibfnamefont {David~I.}\
  \bibnamefont {Kaiser}},\ }\bibfield  {title} {\enquote {\bibinfo {title}
  {{Debye screening of non-Abelian plasmas in curved spacetimes}},}\ }\href
  {\doibase 10.1103/PhysRevD.108.125010} {\bibfield  {journal} {\bibinfo
  {journal} {Phys. Rev. D}\ }\textbf {\bibinfo {volume} {108}},\ \bibinfo
  {pages} {125010} (\bibinfo {year} {2023})},\ \Eprint
  {http://arxiv.org/abs/2309.15385} {arXiv:2309.15385 [hep-ph]} \BibitemShut
  {NoStop}%
\bibitem [{PRL()}]{PRLSM}%
  \BibitemOpen
  \href@noop {} {}\bibinfo {note} {See the Supplemental Material (SM), which
  includes
  Refs.~\cite{ShapiroTeukolsky1983,Richards:2021zbr,Young:2019yug,DeLuca:2019qsy,DeLuca:2020ioi,Musco:2020jjb,Gow:2020cou,Biagetti:2021eep,Ferrante:2022mui,Gow:2022jfb,DeLuca:2022rfz,DeLuca:2023tun},
  for more information about accretion rates for these primordial black holes
  (PBHs), as well as how to evaluate the expected abundance of PBHs with
  various values of mass and enclosed charge, which depends on the PBH mass
  distribution.}\BibitemShut {Stop}%
\bibitem [{\citenamefont {Pisarski}(1989)}]{Pisarski:1988vd}%
  \BibitemOpen
  \bibfield  {author} {\bibinfo {author} {\bibfnamefont {Robert~D.}\
  \bibnamefont {Pisarski}},\ }\bibfield  {title} {\enquote {\bibinfo {title}
  {{Scattering Amplitudes in Hot Gauge Theories}},}\ }\href {\doibase
  10.1103/PhysRevLett.63.1129} {\bibfield  {journal} {\bibinfo  {journal}
  {Phys. Rev. Lett.}\ }\textbf {\bibinfo {volume} {63}},\ \bibinfo {pages}
  {1129} (\bibinfo {year} {1989})}\BibitemShut {NoStop}%
\bibitem [{\citenamefont {Bazavov}\ and\ \citenamefont
  {Weber}(2021)}]{Bazavov:2020teh}%
  \BibitemOpen
  \bibfield  {author} {\bibinfo {author} {\bibfnamefont {Alexei}\ \bibnamefont
  {Bazavov}}\ and\ \bibinfo {author} {\bibfnamefont {Johannes~Heinrich}\
  \bibnamefont {Weber}},\ }\bibfield  {title} {\enquote {\bibinfo {title}
  {{Color screening in quantum chromodynamics}},}\ }\href {\doibase
  10.1016/j.ppnp.2020.103823} {\bibfield  {journal} {\bibinfo  {journal} {Prog.
  Part. Nucl. Phys.}\ }\textbf {\bibinfo {volume} {116}},\ \bibinfo {pages}
  {103823} (\bibinfo {year} {2021})},\ \Eprint
  {http://arxiv.org/abs/2010.01873} {arXiv:2010.01873 [hep-lat]} \BibitemShut
  {NoStop}%
\bibitem [{Tem()}]{TempNote}%
  \BibitemOpen
  \href@noop {} {}\bibinfo {note} {We have neglected the modest temperature
  gradients that will develop within the collapsing fluid, which in turn will
  affect the local Debye screening length. During collapse, the local fluid
  density will rise, which will increase the local temperature and hence
  decrease $\lambda_D (T)$. As shown in Ref.~\cite{Alonso-Monsalve:2023jfq},
  within the hot plasma such thermal gradients are given by the familiar Tolman
  temperature, $T (r) = T_\infty \sqrt{ - g^{00} (r)}$, where $T_\infty$ is the
  temperature of the plasma far from the origin
  \cite{Tolman:1930zza,Tolman:1930ona,Santiago:2018lcy}. Meanwhile, numerical
  simulations of PBH formation typically find a local underdensity at radii $r
  \gtrsim r_c$, where $r_c$ is the collapse radius. (See, e.g.,
  Refs.~\cite{Evans:1994pj,Bloomfield:2015ila,deJong:2021bbo}.) Hence we may
  approximate the local region of spacetime with a line-element in which
  $g_{0i} = 0$ and $g_{00} = 1/g^{00} = -(1-2G M(r) / r)$. Using $M(r_c) / r_c
  \simeq \bar{M} (\bar{r}_c) / \bar{r}_c \simeq \gamma / G$ (following the
  discussion above Eq.~(\ref{MtobarM})), this yields $T (r_c) \simeq 1.3 \,
  T_\infty$, an estimate consistent with the results for the evolving metric
  coefficient $g_{rr}$ shown in Fig.~4 of Ref.~\cite{Evans:1994pj}, since we
  expect $-g^{00} \simeq g_{rr}$ for $0 < r \leq r_c$ during the collapse
  phase, given the local underdensity for $r \gtrsim r_c$ \cite{Visser:1992qh}.
  The effect from $T(r_c)$ on our estimate of $N_{\rm cc}$ in Eq.~(\ref{Ncc})
  would be more than compensated by other effects we have neglected, such as
  the nonnegligible probability that two or more net-color regions with the
  same enclosed charge (e.g., $a = 1$) happened to be contiguous at the origin,
  which would raise ${\cal Q}_0^a$.}\BibitemShut {Stop}%
\bibitem [{\citenamefont {Ade}\ \emph {et~al.}(2021)\citenamefont {Ade} \emph
  {et~al.}}]{BICEP:2021xfz}%
  \BibitemOpen
  \bibfield  {author} {\bibinfo {author} {\bibfnamefont {P.~A.~R.}\
  \bibnamefont {Ade}} \emph {et~al.} (\bibinfo {collaboration} {BICEP, Keck}),\
  }\bibfield  {title} {\enquote {\bibinfo {title} {{Improved Constraints on
  primordial gravitational waves using Planck, WMAP, and BICEP/Keck
  observations through the 2018 observing season}},}\ }\href {\doibase
  10.1103/PhysRevLett.127.151301} {\bibfield  {journal} {\bibinfo  {journal}
  {Phys. Rev. Lett.}\ }\textbf {\bibinfo {volume} {127}},\ \bibinfo {pages}
  {151301} (\bibinfo {year} {2021})},\ \Eprint
  {http://arxiv.org/abs/2110.00483} {arXiv:2110.00483 [astro-ph.CO]}
  \BibitemShut {NoStop}%
\bibitem [{Sta()}]{StabilityNote}%
  \BibitemOpen
  \href@noop {} {}\bibinfo {note} {Although vacuum solutions of charged
  Einstein-Yang-Mills black holes in asymptotically flat spacetime are known to
  be unstable to linear perturbations
  \cite{Straumann:1989tf,Straumann:1990as,Bizon:1991nt,Galtsov:1991nk}, such
  black holes embedded in asymptotically de Sitter spacetime are considerably
  more stable \cite{Torii:1995wv}. It remains to study the stability under
  linearized perturbations of a color-charged PBH immersed in a hot, screening
  medium.}\BibitemShut {Stop}%
\bibitem [{\citenamefont {Feng}\ \emph {et~al.}(2023)\citenamefont {Feng},
  \citenamefont {Chakraborty},\ and\ \citenamefont {Cardoso}}]{Feng:2022evy}%
  \BibitemOpen
  \bibfield  {author} {\bibinfo {author} {\bibfnamefont {Justin~C.}\
  \bibnamefont {Feng}}, \bibinfo {author} {\bibfnamefont {Sumanta}\
  \bibnamefont {Chakraborty}}, \ and\ \bibinfo {author} {\bibfnamefont {Vitor}\
  \bibnamefont {Cardoso}},\ }\bibfield  {title} {\enquote {\bibinfo {title}
  {{Shielding a charged black hole}},}\ }\href {\doibase
  10.1103/PhysRevD.107.044050} {\bibfield  {journal} {\bibinfo  {journal}
  {Phys. Rev. D}\ }\textbf {\bibinfo {volume} {107}},\ \bibinfo {pages}
  {044050} (\bibinfo {year} {2023})},\ \Eprint
  {http://arxiv.org/abs/2211.05261} {arXiv:2211.05261 [gr-qc]} \BibitemShut
  {NoStop}%
\bibitem [{\citenamefont {Dom\`enech}(2021)}]{Domenech:2021ztg}%
  \BibitemOpen
  \bibfield  {author} {\bibinfo {author} {\bibfnamefont {Guillem}\ \bibnamefont
  {Dom\`enech}},\ }\bibfield  {title} {\enquote {\bibinfo {title} {{Scalar
  Induced Gravitational Waves Review}},}\ }\href {\doibase
  10.3390/universe7110398} {\bibfield  {journal} {\bibinfo  {journal}
  {Universe}\ }\textbf {\bibinfo {volume} {7}},\ \bibinfo {pages} {398}
  (\bibinfo {year} {2021})},\ \Eprint {http://arxiv.org/abs/2109.01398}
  {arXiv:2109.01398 [gr-qc]} \BibitemShut {NoStop}%
\bibitem [{\citenamefont {Kawasaki}\ \emph {et~al.}(2005)\citenamefont
  {Kawasaki}, \citenamefont {Kohri},\ and\ \citenamefont
  {Moroi}}]{Kawasaki:2004qu}%
  \BibitemOpen
  \bibfield  {author} {\bibinfo {author} {\bibfnamefont {Masahiro}\
  \bibnamefont {Kawasaki}}, \bibinfo {author} {\bibfnamefont {Kazunori}\
  \bibnamefont {Kohri}}, \ and\ \bibinfo {author} {\bibfnamefont {Takeo}\
  \bibnamefont {Moroi}},\ }\bibfield  {title} {\enquote {\bibinfo {title}
  {{Big-Bang nucleosynthesis and hadronic decay of long-lived massive
  particles}},}\ }\href {\doibase 10.1103/PhysRevD.71.083502} {\bibfield
  {journal} {\bibinfo  {journal} {Phys. Rev. D}\ }\textbf {\bibinfo {volume}
  {71}},\ \bibinfo {pages} {083502} (\bibinfo {year} {2005})},\ \Eprint
  {http://arxiv.org/abs/astro-ph/0408426} {arXiv:astro-ph/0408426} \BibitemShut
  {NoStop}%
\bibitem [{\citenamefont {Carr}\ \emph {et~al.}(2010)\citenamefont {Carr},
  \citenamefont {Kohri}, \citenamefont {Sendouda},\ and\ \citenamefont
  {Yokoyama}}]{Carr:2009jm}%
  \BibitemOpen
  \bibfield  {author} {\bibinfo {author} {\bibfnamefont {B.~J.}\ \bibnamefont
  {Carr}}, \bibinfo {author} {\bibfnamefont {Kazunori}\ \bibnamefont {Kohri}},
  \bibinfo {author} {\bibfnamefont {Yuuiti}\ \bibnamefont {Sendouda}}, \ and\
  \bibinfo {author} {\bibfnamefont {Jun'ichi}\ \bibnamefont {Yokoyama}},\
  }\bibfield  {title} {\enquote {\bibinfo {title} {{New cosmological
  constraints on primordial black holes}},}\ }\href {\doibase
  10.1103/PhysRevD.81.104019} {\bibfield  {journal} {\bibinfo  {journal} {Phys.
  Rev. D}\ }\textbf {\bibinfo {volume} {81}},\ \bibinfo {pages} {104019}
  (\bibinfo {year} {2010})},\ \Eprint {http://arxiv.org/abs/0912.5297}
  {arXiv:0912.5297 [astro-ph.CO]} \BibitemShut {NoStop}%
\bibitem [{\citenamefont {de~Freitas~Pacheco}\ \emph
  {et~al.}(2023)\citenamefont {de~Freitas~Pacheco}, \citenamefont {Kiritsis},
  \citenamefont {Lucca},\ and\ \citenamefont
  {Silk}}]{deFreitasPacheco:2023hpb}%
  \BibitemOpen
  \bibfield  {author} {\bibinfo {author} {\bibfnamefont {Jose~A.}\ \bibnamefont
  {de~Freitas~Pacheco}}, \bibinfo {author} {\bibfnamefont {Elias}\ \bibnamefont
  {Kiritsis}}, \bibinfo {author} {\bibfnamefont {Matteo}\ \bibnamefont
  {Lucca}}, \ and\ \bibinfo {author} {\bibfnamefont {Joseph}\ \bibnamefont
  {Silk}},\ }\bibfield  {title} {\enquote {\bibinfo {title} {{Quasiextremal
  primordial black holes are a viable dark matter candidate}},}\ }\href
  {\doibase 10.1103/PhysRevD.107.123525} {\bibfield  {journal} {\bibinfo
  {journal} {Phys. Rev. D}\ }\textbf {\bibinfo {volume} {107}},\ \bibinfo
  {pages} {123525} (\bibinfo {year} {2023})},\ \Eprint
  {http://arxiv.org/abs/2301.13215} {arXiv:2301.13215 [astro-ph.CO]}
  \BibitemShut {NoStop}%
\bibitem [{\citenamefont {Coleman}\ \emph
  {et~al.}(1991{\natexlab{a}})\citenamefont {Coleman}, \citenamefont
  {Preskill},\ and\ \citenamefont {Wilczek}}]{Coleman:1991sj}%
  \BibitemOpen
  \bibfield  {author} {\bibinfo {author} {\bibfnamefont {Sidney~R.}\
  \bibnamefont {Coleman}}, \bibinfo {author} {\bibfnamefont {John}\
  \bibnamefont {Preskill}}, \ and\ \bibinfo {author} {\bibfnamefont {Frank}\
  \bibnamefont {Wilczek}},\ }\bibfield  {title} {\enquote {\bibinfo {title}
  {{Dynamical effect of quantum hair}},}\ }\href {\doibase
  10.1142/S0217732391001767} {\bibfield  {journal} {\bibinfo  {journal} {Mod.
  Phys. Lett. A}\ }\textbf {\bibinfo {volume} {6}},\ \bibinfo {pages}
  {1631--1642} (\bibinfo {year} {1991}{\natexlab{a}})}\BibitemShut {NoStop}%
\bibitem [{\citenamefont {Coleman}\ \emph
  {et~al.}(1991{\natexlab{b}})\citenamefont {Coleman}, \citenamefont
  {Preskill},\ and\ \citenamefont {Wilczek}}]{Coleman:1991jf}%
  \BibitemOpen
  \bibfield  {author} {\bibinfo {author} {\bibfnamefont {Sidney~R.}\
  \bibnamefont {Coleman}}, \bibinfo {author} {\bibfnamefont {John}\
  \bibnamefont {Preskill}}, \ and\ \bibinfo {author} {\bibfnamefont {Frank}\
  \bibnamefont {Wilczek}},\ }\bibfield  {title} {\enquote {\bibinfo {title}
  {{Growing hair on black holes}},}\ }\href {\doibase
  10.1103/PhysRevLett.67.1975} {\bibfield  {journal} {\bibinfo  {journal}
  {Phys. Rev. Lett.}\ }\textbf {\bibinfo {volume} {67}},\ \bibinfo {pages}
  {1975--1978} (\bibinfo {year} {1991}{\natexlab{b}})}\BibitemShut {NoStop}%
\bibitem [{\citenamefont {Coleman}\ \emph
  {et~al.}(1992{\natexlab{a}})\citenamefont {Coleman}, \citenamefont
  {Preskill},\ and\ \citenamefont {Wilczek}}]{Coleman:1991ku}%
  \BibitemOpen
  \bibfield  {author} {\bibinfo {author} {\bibfnamefont {Sidney~R.}\
  \bibnamefont {Coleman}}, \bibinfo {author} {\bibfnamefont {John}\
  \bibnamefont {Preskill}}, \ and\ \bibinfo {author} {\bibfnamefont {Frank}\
  \bibnamefont {Wilczek}},\ }\bibfield  {title} {\enquote {\bibinfo {title}
  {{Quantum hair on black holes}},}\ }\href {\doibase
  10.1016/0550-3213(92)90008-Y} {\bibfield  {journal} {\bibinfo  {journal}
  {Nucl. Phys. B}\ }\textbf {\bibinfo {volume} {378}},\ \bibinfo {pages}
  {175--246} (\bibinfo {year} {1992}{\natexlab{a}})},\ \Eprint
  {http://arxiv.org/abs/hep-th/9201059} {arXiv:hep-th/9201059} \BibitemShut
  {NoStop}%
\bibitem [{\citenamefont {Coleman}\ \emph
  {et~al.}(1992{\natexlab{b}})\citenamefont {Coleman}, \citenamefont {Krauss},
  \citenamefont {Preskill},\ and\ \citenamefont {Wilczek}}]{Coleman:1992rn}%
  \BibitemOpen
  \bibfield  {author} {\bibinfo {author} {\bibfnamefont {Sidney~R.}\
  \bibnamefont {Coleman}}, \bibinfo {author} {\bibfnamefont {L.~M.}\
  \bibnamefont {Krauss}}, \bibinfo {author} {\bibfnamefont {John}\ \bibnamefont
  {Preskill}}, \ and\ \bibinfo {author} {\bibfnamefont {Frank}\ \bibnamefont
  {Wilczek}},\ }\bibfield  {title} {\enquote {\bibinfo {title} {{Quantum hair
  and quantum gravity}},}\ }\href {\doibase 10.1007/BF00756870} {\bibfield
  {journal} {\bibinfo  {journal} {Gen. Rel. Grav.}\ }\textbf {\bibinfo {volume}
  {24}},\ \bibinfo {pages} {9--16} (\bibinfo {year}
  {1992}{\natexlab{b}})}\BibitemShut {NoStop}%
\bibitem [{\citenamefont {Krauss}\ and\ \citenamefont
  {Liu}(1997)}]{Krauss:1996df}%
  \BibitemOpen
  \bibfield  {author} {\bibinfo {author} {\bibfnamefont {Lawrence~M.}\
  \bibnamefont {Krauss}}\ and\ \bibinfo {author} {\bibfnamefont {Hong}\
  \bibnamefont {Liu}},\ }\bibfield  {title} {\enquote {\bibinfo {title}
  {{Quantum hair, instantons, and black hole thermodynamics: Some new
  results}},}\ }\href {\doibase 10.1016/S0550-3213(97)00053-9} {\bibfield
  {journal} {\bibinfo  {journal} {Nucl. Phys. B}\ }\textbf {\bibinfo {volume}
  {491}},\ \bibinfo {pages} {365--386} (\bibinfo {year} {1997})},\ \Eprint
  {http://arxiv.org/abs/hep-th/9611032} {arXiv:hep-th/9611032} \BibitemShut
  {NoStop}%
\bibitem [{\citenamefont {Garcia~Garcia}(2019)}]{GarciaGarcia:2018tua}%
  \BibitemOpen
  \bibfield  {author} {\bibinfo {author} {\bibfnamefont {Isabel}\ \bibnamefont
  {Garcia~Garcia}},\ }\bibfield  {title} {\enquote {\bibinfo {title}
  {{Properties of Discrete Black Hole Hair}},}\ }\href {\doibase
  10.1007/JHEP02(2019)117} {\bibfield  {journal} {\bibinfo  {journal} {JHEP}\
  }\textbf {\bibinfo {volume} {02}},\ \bibinfo {pages} {117} (\bibinfo {year}
  {2019})},\ \Eprint {http://arxiv.org/abs/1809.03527} {arXiv:1809.03527
  [gr-qc]} \BibitemShut {NoStop}%
\bibitem [{\citenamefont {Horowitz}\ \emph {et~al.}(2016)\citenamefont
  {Horowitz}, \citenamefont {Santos},\ and\ \citenamefont
  {Way}}]{Horowitz:2016ezu}%
  \BibitemOpen
  \bibfield  {author} {\bibinfo {author} {\bibfnamefont {Gary~T.}\ \bibnamefont
  {Horowitz}}, \bibinfo {author} {\bibfnamefont {Jorge~E.}\ \bibnamefont
  {Santos}}, \ and\ \bibinfo {author} {\bibfnamefont {Benson}\ \bibnamefont
  {Way}},\ }\bibfield  {title} {\enquote {\bibinfo {title} {{Evidence for an
  Electrifying Violation of Cosmic Censorship}},}\ }\href {\doibase
  10.1088/0264-9381/33/19/195007} {\bibfield  {journal} {\bibinfo  {journal}
  {Class. Quant. Grav.}\ }\textbf {\bibinfo {volume} {33}},\ \bibinfo {pages}
  {195007} (\bibinfo {year} {2016})},\ \Eprint
  {http://arxiv.org/abs/1604.06465} {arXiv:1604.06465 [hep-th]} \BibitemShut
  {NoStop}%
\bibitem [{\citenamefont {Crisford}\ \emph {et~al.}(2018)\citenamefont
  {Crisford}, \citenamefont {Horowitz},\ and\ \citenamefont
  {Santos}}]{Crisford:2017gsb}%
  \BibitemOpen
  \bibfield  {author} {\bibinfo {author} {\bibfnamefont {Toby}\ \bibnamefont
  {Crisford}}, \bibinfo {author} {\bibfnamefont {Gary~T.}\ \bibnamefont
  {Horowitz}}, \ and\ \bibinfo {author} {\bibfnamefont {Jorge~E.}\ \bibnamefont
  {Santos}},\ }\bibfield  {title} {\enquote {\bibinfo {title} {{Testing the
  Weak Gravity - Cosmic Censorship Connection}},}\ }\href {\doibase
  10.1103/PhysRevD.97.066005} {\bibfield  {journal} {\bibinfo  {journal} {Phys.
  Rev. D}\ }\textbf {\bibinfo {volume} {97}},\ \bibinfo {pages} {066005}
  (\bibinfo {year} {2018})},\ \Eprint {http://arxiv.org/abs/1709.07880}
  {arXiv:1709.07880 [hep-th]} \BibitemShut {NoStop}%
\bibitem [{\citenamefont {Horowitz}\ and\ \citenamefont
  {Santos}(2019)}]{Horowitz:2019eum}%
  \BibitemOpen
  \bibfield  {author} {\bibinfo {author} {\bibfnamefont {Gary~T.}\ \bibnamefont
  {Horowitz}}\ and\ \bibinfo {author} {\bibfnamefont {Jorge~E.}\ \bibnamefont
  {Santos}},\ }\bibfield  {title} {\enquote {\bibinfo {title} {{Further
  evidence for the weak gravity \textemdash{} cosmic censorship connection}},}\
  }\href {\doibase 10.1007/JHEP06(2019)122} {\bibfield  {journal} {\bibinfo
  {journal} {JHEP}\ }\textbf {\bibinfo {volume} {06}},\ \bibinfo {pages} {122}
  (\bibinfo {year} {2019})},\ \Eprint {http://arxiv.org/abs/1901.11096}
  {arXiv:1901.11096 [hep-th]} \BibitemShut {NoStop}%
\bibitem [{\citenamefont {Engelhardt}\ and\ \citenamefont
  {Horowitz}(2019)}]{Engelhardt:2019btp}%
  \BibitemOpen
  \bibfield  {author} {\bibinfo {author} {\bibfnamefont {Netta}\ \bibnamefont
  {Engelhardt}}\ and\ \bibinfo {author} {\bibfnamefont {Gary~T.}\ \bibnamefont
  {Horowitz}},\ }\bibfield  {title} {\enquote {\bibinfo {title} {{Holographic
  argument for the Penrose inequality in AdS spacetimes}},}\ }\href {\doibase
  10.1103/PhysRevD.99.126009} {\bibfield  {journal} {\bibinfo  {journal} {Phys.
  Rev. D}\ }\textbf {\bibinfo {volume} {99}},\ \bibinfo {pages} {126009}
  (\bibinfo {year} {2019})},\ \Eprint {http://arxiv.org/abs/1903.00555}
  {arXiv:1903.00555 [hep-th]} \BibitemShut {NoStop}%
\bibitem [{\citenamefont {Kopp}\ \emph {et~al.}(2011)\citenamefont {Kopp},
  \citenamefont {Hofmann},\ and\ \citenamefont {Weller}}]{Kopp:2010sh}%
  \BibitemOpen
  \bibfield  {author} {\bibinfo {author} {\bibfnamefont {Michael}\ \bibnamefont
  {Kopp}}, \bibinfo {author} {\bibfnamefont {Stefan}\ \bibnamefont {Hofmann}},
  \ and\ \bibinfo {author} {\bibfnamefont {Jochen}\ \bibnamefont {Weller}},\
  }\bibfield  {title} {\enquote {\bibinfo {title} {{Separate Universes Do Not
  Constrain Primordial Black Hole Formation}},}\ }\href {\doibase
  10.1103/PhysRevD.83.124025} {\bibfield  {journal} {\bibinfo  {journal} {Phys.
  Rev. D}\ }\textbf {\bibinfo {volume} {83}},\ \bibinfo {pages} {124025}
  (\bibinfo {year} {2011})},\ \Eprint {http://arxiv.org/abs/1012.4369}
  {arXiv:1012.4369 [astro-ph.CO]} \BibitemShut {NoStop}%
\bibitem [{\citenamefont {Carr}\ and\ \citenamefont
  {Harada}(2015)}]{Carr:2014pga}%
  \BibitemOpen
  \bibfield  {author} {\bibinfo {author} {\bibfnamefont {B.~J.}\ \bibnamefont
  {Carr}}\ and\ \bibinfo {author} {\bibfnamefont {Tomohiro}\ \bibnamefont
  {Harada}},\ }\bibfield  {title} {\enquote {\bibinfo {title} {{Separate
  universe problem: 40 years on}},}\ }\href {\doibase
  10.1103/PhysRevD.91.084048} {\bibfield  {journal} {\bibinfo  {journal} {Phys.
  Rev. D}\ }\textbf {\bibinfo {volume} {91}},\ \bibinfo {pages} {084048}
  (\bibinfo {year} {2015})},\ \Eprint {http://arxiv.org/abs/1405.3624}
  {arXiv:1405.3624 [astro-ph.CO]} \BibitemShut {NoStop}%
\bibitem [{\citenamefont {Uehara}\ \emph {et~al.}(2024)\citenamefont {Uehara},
  \citenamefont {Escriv\`a}, \citenamefont {Harada}, \citenamefont {Saito},\
  and\ \citenamefont {Yoo}}]{Uehara:2024yyp}%
  \BibitemOpen
  \bibfield  {author} {\bibinfo {author} {\bibfnamefont {Koichiro}\
  \bibnamefont {Uehara}}, \bibinfo {author} {\bibfnamefont {Albert}\
  \bibnamefont {Escriv\`a}}, \bibinfo {author} {\bibfnamefont {Tomohiro}\
  \bibnamefont {Harada}}, \bibinfo {author} {\bibfnamefont {Daiki}\
  \bibnamefont {Saito}}, \ and\ \bibinfo {author} {\bibfnamefont {Chul-Moon}\
  \bibnamefont {Yoo}},\ }\bibfield  {title} {\enquote {\bibinfo {title}
  {{Numerical simulation of type II primordial black hole formation}},}\
  }\href@noop {} {\  (\bibinfo {year} {2024})},\ \Eprint
  {http://arxiv.org/abs/2401.06329} {arXiv:2401.06329 [gr-qc]} \BibitemShut
  {NoStop}%
\bibitem [{\citenamefont {{Shapiro}}\ and\ \citenamefont
  {{Teukolsky}}(1983)}]{ShapiroTeukolsky1983}%
  \BibitemOpen
  \bibfield  {author} {\bibinfo {author} {\bibfnamefont {Stuart~L.}\
  \bibnamefont {{Shapiro}}}\ and\ \bibinfo {author} {\bibfnamefont {Saul~A.}\
  \bibnamefont {{Teukolsky}}},\ }\href {\doibase 10.1002/9783527617661} {\emph
  {\bibinfo {title} {{Black Holes, White Dwarfs, and Neutron Stars: The Physics
  of Compact Objects}}}}\ (\bibinfo  {publisher} {Wiley},\ \bibinfo {address}
  {New York},\ \bibinfo {year} {1983})\BibitemShut {NoStop}%
\bibitem [{\citenamefont {Richards}\ \emph {et~al.}(2021)\citenamefont
  {Richards}, \citenamefont {Baumgarte},\ and\ \citenamefont
  {Shapiro}}]{Richards:2021zbr}%
  \BibitemOpen
  \bibfield  {author} {\bibinfo {author} {\bibfnamefont {Chloe~B.}\
  \bibnamefont {Richards}}, \bibinfo {author} {\bibfnamefont {Thomas~W.}\
  \bibnamefont {Baumgarte}}, \ and\ \bibinfo {author} {\bibfnamefont
  {Stuart~L.}\ \bibnamefont {Shapiro}},\ }\bibfield  {title} {\enquote
  {\bibinfo {title} {{Relativistic Bondi accretion for stiff equations of
  state}},}\ }\href {\doibase 10.1093/mnras/stab2069} {\bibfield  {journal}
  {\bibinfo  {journal} {Mon. Not. Roy. Astron. Soc.}\ }\textbf {\bibinfo
  {volume} {502}},\ \bibinfo {pages} {3003--3011} (\bibinfo {year} {2021})},\
  \bibinfo {note} {[Erratum: Mon.Not.Roy.Astron.Soc. 506, 3935 (2021)]},\
  \Eprint {http://arxiv.org/abs/2101.08797} {arXiv:2101.08797 [astro-ph.HE]}
  \BibitemShut {NoStop}%
\bibitem [{\citenamefont {Young}\ \emph {et~al.}(2019)\citenamefont {Young},
  \citenamefont {Musco},\ and\ \citenamefont {Byrnes}}]{Young:2019yug}%
  \BibitemOpen
  \bibfield  {author} {\bibinfo {author} {\bibfnamefont {Sam}\ \bibnamefont
  {Young}}, \bibinfo {author} {\bibfnamefont {Ilia}\ \bibnamefont {Musco}}, \
  and\ \bibinfo {author} {\bibfnamefont {Christian~T.}\ \bibnamefont
  {Byrnes}},\ }\bibfield  {title} {\enquote {\bibinfo {title} {{Primordial
  black hole formation and abundance: contribution from the non-linear relation
  between the density and curvature perturbation}},}\ }\href {\doibase
  10.1088/1475-7516/2019/11/012} {\bibfield  {journal} {\bibinfo  {journal}
  {JCAP}\ }\textbf {\bibinfo {volume} {11}},\ \bibinfo {pages} {012} (\bibinfo
  {year} {2019})},\ \Eprint {http://arxiv.org/abs/1904.00984} {arXiv:1904.00984
  [astro-ph.CO]} \BibitemShut {NoStop}%
\bibitem [{\citenamefont {De~Luca}\ \emph {et~al.}(2019)\citenamefont
  {De~Luca}, \citenamefont {Franciolini}, \citenamefont {Kehagias},
  \citenamefont {Peloso}, \citenamefont {Riotto},\ and\ \citenamefont
  {\"Unal}}]{DeLuca:2019qsy}%
  \BibitemOpen
  \bibfield  {author} {\bibinfo {author} {\bibfnamefont {V.}~\bibnamefont
  {De~Luca}}, \bibinfo {author} {\bibfnamefont {G.}~\bibnamefont
  {Franciolini}}, \bibinfo {author} {\bibfnamefont {A.}~\bibnamefont
  {Kehagias}}, \bibinfo {author} {\bibfnamefont {M.}~\bibnamefont {Peloso}},
  \bibinfo {author} {\bibfnamefont {A.}~\bibnamefont {Riotto}}, \ and\ \bibinfo
  {author} {\bibfnamefont {C.}~\bibnamefont {\"Unal}},\ }\bibfield  {title}
  {\enquote {\bibinfo {title} {{The Ineludible non-Gaussianity of the
  Primordial Black Hole Abundance}},}\ }\href {\doibase
  10.1088/1475-7516/2019/07/048} {\bibfield  {journal} {\bibinfo  {journal}
  {JCAP}\ }\textbf {\bibinfo {volume} {07}},\ \bibinfo {pages} {048} (\bibinfo
  {year} {2019})},\ \Eprint {http://arxiv.org/abs/1904.00970} {arXiv:1904.00970
  [astro-ph.CO]} \BibitemShut {NoStop}%
\bibitem [{\citenamefont {De~Luca}\ \emph
  {et~al.}(2020{\natexlab{b}})\citenamefont {De~Luca}, \citenamefont
  {Franciolini},\ and\ \citenamefont {Riotto}}]{DeLuca:2020ioi}%
  \BibitemOpen
  \bibfield  {author} {\bibinfo {author} {\bibfnamefont {V.}~\bibnamefont
  {De~Luca}}, \bibinfo {author} {\bibfnamefont {G.}~\bibnamefont
  {Franciolini}}, \ and\ \bibinfo {author} {\bibfnamefont {A.}~\bibnamefont
  {Riotto}},\ }\bibfield  {title} {\enquote {\bibinfo {title} {{On the
  Primordial Black Hole Mass Function for Broad Spectra}},}\ }\href {\doibase
  10.1016/j.physletb.2020.135550} {\bibfield  {journal} {\bibinfo  {journal}
  {Phys. Lett. B}\ }\textbf {\bibinfo {volume} {807}},\ \bibinfo {pages}
  {135550} (\bibinfo {year} {2020}{\natexlab{b}})},\ \Eprint
  {http://arxiv.org/abs/2001.04371} {arXiv:2001.04371 [astro-ph.CO]}
  \BibitemShut {NoStop}%
\bibitem [{\citenamefont {Musco}\ \emph {et~al.}(2021)\citenamefont {Musco},
  \citenamefont {De~Luca}, \citenamefont {Franciolini},\ and\ \citenamefont
  {Riotto}}]{Musco:2020jjb}%
  \BibitemOpen
  \bibfield  {author} {\bibinfo {author} {\bibfnamefont {Ilia}\ \bibnamefont
  {Musco}}, \bibinfo {author} {\bibfnamefont {Valerio}\ \bibnamefont
  {De~Luca}}, \bibinfo {author} {\bibfnamefont {Gabriele}\ \bibnamefont
  {Franciolini}}, \ and\ \bibinfo {author} {\bibfnamefont {Antonio}\
  \bibnamefont {Riotto}},\ }\bibfield  {title} {\enquote {\bibinfo {title}
  {{Threshold for primordial black holes. II. A simple analytic
  prescription}},}\ }\href {\doibase 10.1103/PhysRevD.103.063538} {\bibfield
  {journal} {\bibinfo  {journal} {Phys. Rev. D}\ }\textbf {\bibinfo {volume}
  {103}},\ \bibinfo {pages} {063538} (\bibinfo {year} {2021})},\ \Eprint
  {http://arxiv.org/abs/2011.03014} {arXiv:2011.03014 [astro-ph.CO]}
  \BibitemShut {NoStop}%
\bibitem [{\citenamefont {Gow}\ \emph {et~al.}(2022)\citenamefont {Gow},
  \citenamefont {Byrnes},\ and\ \citenamefont {Hall}}]{Gow:2020cou}%
  \BibitemOpen
  \bibfield  {author} {\bibinfo {author} {\bibfnamefont {Andrew~D.}\
  \bibnamefont {Gow}}, \bibinfo {author} {\bibfnamefont {Christian~T.}\
  \bibnamefont {Byrnes}}, \ and\ \bibinfo {author} {\bibfnamefont {Alex}\
  \bibnamefont {Hall}},\ }\bibfield  {title} {\enquote {\bibinfo {title}
  {{Accurate model for the primordial black hole mass distribution from a peak
  in the power spectrum}},}\ }\href {\doibase 10.1103/PhysRevD.105.023503}
  {\bibfield  {journal} {\bibinfo  {journal} {Phys. Rev. D}\ }\textbf {\bibinfo
  {volume} {105}},\ \bibinfo {pages} {023503} (\bibinfo {year} {2022})},\
  \Eprint {http://arxiv.org/abs/2009.03204} {arXiv:2009.03204 [astro-ph.CO]}
  \BibitemShut {NoStop}%
\bibitem [{\citenamefont {Biagetti}\ \emph {et~al.}(2021)\citenamefont
  {Biagetti}, \citenamefont {De~Luca}, \citenamefont {Franciolini},
  \citenamefont {Kehagias},\ and\ \citenamefont {Riotto}}]{Biagetti:2021eep}%
  \BibitemOpen
  \bibfield  {author} {\bibinfo {author} {\bibfnamefont {Matteo}\ \bibnamefont
  {Biagetti}}, \bibinfo {author} {\bibfnamefont {Valerio}\ \bibnamefont
  {De~Luca}}, \bibinfo {author} {\bibfnamefont {Gabriele}\ \bibnamefont
  {Franciolini}}, \bibinfo {author} {\bibfnamefont {Alex}\ \bibnamefont
  {Kehagias}}, \ and\ \bibinfo {author} {\bibfnamefont {Antonio}\ \bibnamefont
  {Riotto}},\ }\bibfield  {title} {\enquote {\bibinfo {title} {{The formation
  probability of primordial black holes}},}\ }\href {\doibase
  10.1016/j.physletb.2021.136602} {\bibfield  {journal} {\bibinfo  {journal}
  {Phys. Lett. B}\ }\textbf {\bibinfo {volume} {820}},\ \bibinfo {pages}
  {136602} (\bibinfo {year} {2021})},\ \Eprint
  {http://arxiv.org/abs/2105.07810} {arXiv:2105.07810 [astro-ph.CO]}
  \BibitemShut {NoStop}%
\bibitem [{\citenamefont {Ferrante}\ \emph {et~al.}(2023)\citenamefont
  {Ferrante}, \citenamefont {Franciolini}, \citenamefont {Iovino},\ and\
  \citenamefont {Urbano}}]{Ferrante:2022mui}%
  \BibitemOpen
  \bibfield  {author} {\bibinfo {author} {\bibfnamefont {Giacomo}\ \bibnamefont
  {Ferrante}}, \bibinfo {author} {\bibfnamefont {Gabriele}\ \bibnamefont
  {Franciolini}}, \bibinfo {author} {\bibfnamefont {Antonio}\ \bibnamefont
  {Iovino}, \bibfnamefont {Junior}}, \ and\ \bibinfo {author} {\bibfnamefont
  {Alfredo}\ \bibnamefont {Urbano}},\ }\bibfield  {title} {\enquote {\bibinfo
  {title} {{Primordial non-Gaussianity up to all orders: Theoretical aspects
  and implications for primordial black hole models}},}\ }\href {\doibase
  10.1103/PhysRevD.107.043520} {\bibfield  {journal} {\bibinfo  {journal}
  {Phys. Rev. D}\ }\textbf {\bibinfo {volume} {107}},\ \bibinfo {pages}
  {043520} (\bibinfo {year} {2023})},\ \Eprint
  {http://arxiv.org/abs/2211.01728} {arXiv:2211.01728 [astro-ph.CO]}
  \BibitemShut {NoStop}%
\bibitem [{\citenamefont {Gow}\ \emph {et~al.}(2023)\citenamefont {Gow},
  \citenamefont {Assadullahi}, \citenamefont {Jackson}, \citenamefont {Koyama},
  \citenamefont {Vennin},\ and\ \citenamefont {Wands}}]{Gow:2022jfb}%
  \BibitemOpen
  \bibfield  {author} {\bibinfo {author} {\bibfnamefont {Andrew~D.}\
  \bibnamefont {Gow}}, \bibinfo {author} {\bibfnamefont {Hooshyar}\
  \bibnamefont {Assadullahi}}, \bibinfo {author} {\bibfnamefont {Joseph H.~P.}\
  \bibnamefont {Jackson}}, \bibinfo {author} {\bibfnamefont {Kazuya}\
  \bibnamefont {Koyama}}, \bibinfo {author} {\bibfnamefont {Vincent}\
  \bibnamefont {Vennin}}, \ and\ \bibinfo {author} {\bibfnamefont {David}\
  \bibnamefont {Wands}},\ }\bibfield  {title} {\enquote {\bibinfo {title}
  {{Non-perturbative non-Gaussianity and primordial black holes}},}\ }\href
  {\doibase 10.1209/0295-5075/acd417} {\bibfield  {journal} {\bibinfo
  {journal} {EPL}\ }\textbf {\bibinfo {volume} {142}},\ \bibinfo {pages}
  {49001} (\bibinfo {year} {2023})},\ \Eprint {http://arxiv.org/abs/2211.08348}
  {arXiv:2211.08348 [astro-ph.CO]} \BibitemShut {NoStop}%
\bibitem [{\citenamefont {De~Luca}\ and\ \citenamefont
  {Riotto}(2022)}]{DeLuca:2022rfz}%
  \BibitemOpen
  \bibfield  {author} {\bibinfo {author} {\bibfnamefont {V.}~\bibnamefont
  {De~Luca}}\ and\ \bibinfo {author} {\bibfnamefont {A.}~\bibnamefont
  {Riotto}},\ }\bibfield  {title} {\enquote {\bibinfo {title} {{A note on the
  abundance of primordial black holes: Use and misuse of the metric curvature
  perturbation}},}\ }\href {\doibase 10.1016/j.physletb.2022.137035} {\bibfield
   {journal} {\bibinfo  {journal} {Phys. Lett. B}\ }\textbf {\bibinfo {volume}
  {828}},\ \bibinfo {pages} {137035} (\bibinfo {year} {2022})},\ \Eprint
  {http://arxiv.org/abs/2201.09008} {arXiv:2201.09008 [astro-ph.CO]}
  \BibitemShut {NoStop}%
\bibitem [{\citenamefont {De~Luca}\ \emph {et~al.}(2023)\citenamefont
  {De~Luca}, \citenamefont {Kehagias},\ and\ \citenamefont
  {Riotto}}]{DeLuca:2023tun}%
  \BibitemOpen
  \bibfield  {author} {\bibinfo {author} {\bibfnamefont {Valerio}\ \bibnamefont
  {De~Luca}}, \bibinfo {author} {\bibfnamefont {Alex}\ \bibnamefont
  {Kehagias}}, \ and\ \bibinfo {author} {\bibfnamefont {Antonio}\ \bibnamefont
  {Riotto}},\ }\bibfield  {title} {\enquote {\bibinfo {title} {{How well do we
  know the primordial black hole abundance? The crucial role of nonlinearities
  when approaching the horizon}},}\ }\href@noop {} {\  (\bibinfo {year}
  {2023})},\ \Eprint {http://arxiv.org/abs/2307.13633} {arXiv:2307.13633
  [astro-ph.CO]} \BibitemShut {NoStop}%
\bibitem [{\citenamefont {Tolman}(1930)}]{Tolman:1930zza}%
  \BibitemOpen
  \bibfield  {author} {\bibinfo {author} {\bibfnamefont {Richard~C.}\
  \bibnamefont {Tolman}},\ }\bibfield  {title} {\enquote {\bibinfo {title} {{On
  the Weight of Heat and Thermal Equilibrium in General Relativity}},}\ }\href
  {\doibase 10.1103/PhysRev.35.904} {\bibfield  {journal} {\bibinfo  {journal}
  {Phys. Rev.}\ }\textbf {\bibinfo {volume} {35}},\ \bibinfo {pages} {904--924}
  (\bibinfo {year} {1930})}\BibitemShut {NoStop}%
\bibitem [{\citenamefont {Tolman}\ and\ \citenamefont
  {Ehrenfest}(1930)}]{Tolman:1930ona}%
  \BibitemOpen
  \bibfield  {author} {\bibinfo {author} {\bibfnamefont {Richard}\ \bibnamefont
  {Tolman}}\ and\ \bibinfo {author} {\bibfnamefont {Paul}\ \bibnamefont
  {Ehrenfest}},\ }\bibfield  {title} {\enquote {\bibinfo {title} {{Temperature
  Equilibrium in a Static Gravitational Field}},}\ }\href {\doibase
  10.1103/PhysRev.36.1791} {\bibfield  {journal} {\bibinfo  {journal} {Phys.
  Rev.}\ }\textbf {\bibinfo {volume} {36}},\ \bibinfo {pages} {1791--1798}
  (\bibinfo {year} {1930})}\BibitemShut {NoStop}%
\bibitem [{\citenamefont {Santiago}\ and\ \citenamefont
  {Visser}(2018)}]{Santiago:2018lcy}%
  \BibitemOpen
  \bibfield  {author} {\bibinfo {author} {\bibfnamefont {Jessica}\ \bibnamefont
  {Santiago}}\ and\ \bibinfo {author} {\bibfnamefont {Matt}\ \bibnamefont
  {Visser}},\ }\bibfield  {title} {\enquote {\bibinfo {title} {{Tolman-like
  temperature gradients in stationary spacetimes}},}\ }\href {\doibase
  10.1103/PhysRevD.98.064001} {\bibfield  {journal} {\bibinfo  {journal} {Phys.
  Rev. D}\ }\textbf {\bibinfo {volume} {98}},\ \bibinfo {pages} {064001}
  (\bibinfo {year} {2018})},\ \Eprint {http://arxiv.org/abs/1807.02915}
  {arXiv:1807.02915 [gr-qc]} \BibitemShut {NoStop}%
\bibitem [{\citenamefont {Bloomfield}\ \emph {et~al.}(2015)\citenamefont
  {Bloomfield}, \citenamefont {Bulhosa},\ and\ \citenamefont
  {Face}}]{Bloomfield:2015ila}%
  \BibitemOpen
  \bibfield  {author} {\bibinfo {author} {\bibfnamefont {Jolyon}\ \bibnamefont
  {Bloomfield}}, \bibinfo {author} {\bibfnamefont {Daniel}\ \bibnamefont
  {Bulhosa}}, \ and\ \bibinfo {author} {\bibfnamefont {Stephen}\ \bibnamefont
  {Face}},\ }\bibfield  {title} {\enquote {\bibinfo {title} {{Formalism for
  Primordial Black Hole Formation in Spherical Symmetry}},}\ }\href@noop {} {\
  (\bibinfo {year} {2015})},\ \Eprint {http://arxiv.org/abs/1504.02071}
  {arXiv:1504.02071 [gr-qc]} \BibitemShut {NoStop}%
\bibitem [{\citenamefont {de~Jong}\ \emph {et~al.}(2022)\citenamefont
  {de~Jong}, \citenamefont {Aurrekoetxea},\ and\ \citenamefont
  {Lim}}]{deJong:2021bbo}%
  \BibitemOpen
  \bibfield  {author} {\bibinfo {author} {\bibfnamefont {Eloy}\ \bibnamefont
  {de~Jong}}, \bibinfo {author} {\bibfnamefont {Josu~C.}\ \bibnamefont
  {Aurrekoetxea}}, \ and\ \bibinfo {author} {\bibfnamefont {Eugene~A.}\
  \bibnamefont {Lim}},\ }\bibfield  {title} {\enquote {\bibinfo {title}
  {{Primordial black hole formation with full numerical relativity}},}\ }\href
  {\doibase 10.1088/1475-7516/2022/03/029} {\bibfield  {journal} {\bibinfo
  {journal} {JCAP}\ }\textbf {\bibinfo {volume} {03}},\ \bibinfo {pages} {029}
  (\bibinfo {year} {2022})},\ \Eprint {http://arxiv.org/abs/2109.04896}
  {arXiv:2109.04896 [astro-ph.CO]} \BibitemShut {NoStop}%
\bibitem [{\citenamefont {Visser}(1992)}]{Visser:1992qh}%
  \BibitemOpen
  \bibfield  {author} {\bibinfo {author} {\bibfnamefont {Matt}\ \bibnamefont
  {Visser}},\ }\bibfield  {title} {\enquote {\bibinfo {title} {{Dirty black
  holes: Thermodynamics and horizon structure}},}\ }\href {\doibase
  10.1103/PhysRevD.46.2445} {\bibfield  {journal} {\bibinfo  {journal} {Phys.
  Rev. D}\ }\textbf {\bibinfo {volume} {46}},\ \bibinfo {pages} {2445--2451}
  (\bibinfo {year} {1992})},\ \Eprint {http://arxiv.org/abs/hep-th/9203057}
  {arXiv:hep-th/9203057} \BibitemShut {NoStop}%
\bibitem [{\citenamefont {Straumann}\ and\ \citenamefont
  {Zhou}(1990{\natexlab{a}})}]{Straumann:1989tf}%
  \BibitemOpen
  \bibfield  {author} {\bibinfo {author} {\bibfnamefont {Norbert}\ \bibnamefont
  {Straumann}}\ and\ \bibinfo {author} {\bibfnamefont {Zhi-Hong}\ \bibnamefont
  {Zhou}},\ }\bibfield  {title} {\enquote {\bibinfo {title} {{Instability of
  the Bartnik-mckinnon Solution of the Einstein {Yang-Mills} Equations}},}\
  }\href {\doibase 10.1016/0370-2693(90)91188-H} {\bibfield  {journal}
  {\bibinfo  {journal} {Phys. Lett. B}\ }\textbf {\bibinfo {volume} {237}},\
  \bibinfo {pages} {353--356} (\bibinfo {year}
  {1990}{\natexlab{a}})}\BibitemShut {NoStop}%
\bibitem [{\citenamefont {Straumann}\ and\ \citenamefont
  {Zhou}(1990{\natexlab{b}})}]{Straumann:1990as}%
  \BibitemOpen
  \bibfield  {author} {\bibinfo {author} {\bibfnamefont {N.}~\bibnamefont
  {Straumann}}\ and\ \bibinfo {author} {\bibfnamefont {Z.~H.}\ \bibnamefont
  {Zhou}},\ }\bibfield  {title} {\enquote {\bibinfo {title} {{Instability of a
  colored black hole solution}},}\ }\href {\doibase
  10.1016/0370-2693(90)90951-2} {\bibfield  {journal} {\bibinfo  {journal}
  {Phys. Lett. B}\ }\textbf {\bibinfo {volume} {243}},\ \bibinfo {pages}
  {33--35} (\bibinfo {year} {1990}{\natexlab{b}})}\BibitemShut {NoStop}%
\bibitem [{\citenamefont {Bizon}\ and\ \citenamefont
  {Wald}(1991)}]{Bizon:1991nt}%
  \BibitemOpen
  \bibfield  {author} {\bibinfo {author} {\bibfnamefont {P.}~\bibnamefont
  {Bizon}}\ and\ \bibinfo {author} {\bibfnamefont {Robert~M.}\ \bibnamefont
  {Wald}},\ }\bibfield  {title} {\enquote {\bibinfo {title} {{The N=1 colored
  black hole is unstable}},}\ }\href {\doibase 10.1016/0370-2693(91)91243-O}
  {\bibfield  {journal} {\bibinfo  {journal} {Phys. Lett. B}\ }\textbf
  {\bibinfo {volume} {267}},\ \bibinfo {pages} {173--174} (\bibinfo {year}
  {1991})}\BibitemShut {NoStop}%
\bibitem [{\citenamefont {Galtsov}\ and\ \citenamefont
  {Volkov}(1992)}]{Galtsov:1991nk}%
  \BibitemOpen
  \bibfield  {author} {\bibinfo {author} {\bibfnamefont {D.~V.}\ \bibnamefont
  {Galtsov}}\ and\ \bibinfo {author} {\bibfnamefont {M.~S.}\ \bibnamefont
  {Volkov}},\ }\bibfield  {title} {\enquote {\bibinfo {title} {{Instability of
  Einstein Yang-Mills black holes}},}\ }\href {\doibase
  10.1016/0375-9601(92)90990-4} {\bibfield  {journal} {\bibinfo  {journal}
  {Phys. Lett. A}\ }\textbf {\bibinfo {volume} {162}},\ \bibinfo {pages}
  {144--148} (\bibinfo {year} {1992})}\BibitemShut {NoStop}%
\bibitem [{\citenamefont {Torii}\ \emph {et~al.}(1995)\citenamefont {Torii},
  \citenamefont {Maeda},\ and\ \citenamefont {Tachizawa}}]{Torii:1995wv}%
  \BibitemOpen
  \bibfield  {author} {\bibinfo {author} {\bibfnamefont {Takashi}\ \bibnamefont
  {Torii}}, \bibinfo {author} {\bibfnamefont {Kei-ichi}\ \bibnamefont {Maeda}},
  \ and\ \bibinfo {author} {\bibfnamefont {Takashi}\ \bibnamefont
  {Tachizawa}},\ }\bibfield  {title} {\enquote {\bibinfo {title} {{Cosmic
  colored black holes}},}\ }\href {\doibase 10.1103/PhysRevD.52.R4272}
  {\bibfield  {journal} {\bibinfo  {journal} {Phys. Rev. D}\ }\textbf {\bibinfo
  {volume} {52}},\ \bibinfo {pages} {R4272--R4276} (\bibinfo {year} {1995})},\
  \Eprint {http://arxiv.org/abs/gr-qc/9506018} {arXiv:gr-qc/9506018}
  \BibitemShut {NoStop}%
\end{thebibliography}

\end{document}